\documentclass[]{aa} 
 
\usepackage{graphicx, times, txfonts}

\newcommand{\msol}{\mbox{$M_{\odot}$}} 
\newcommand{\rsol}{\mbox{$R_{\odot}$}} 
\newcommand{\msolyr}{{$M_{\odot}$}\,yr$^{-1}$} 
\newcommand{\mdot}{$\dot{M}$}
\newcommand{\lsol}{\mbox{$L_{\odot}$}}

\newcommand{\ks}{km s$^{-1}$}

\begin{document}

\title{
Infrared excess around nearby RGB stars and Reimers law\thanks{
Appendix~A is available in the on-line edition of A\&A. 
} 
}  
 
\author{ 
M.~A.~T.~Groenewegen 
}

\institute{ 
Koninklijke Sterrenwacht van Belgi\"e, Ringlaan 3, B--1180 Brussel, Belgium \\ \email{marting@oma.be}
} 
 
\date{received: 2011,  accepted: 2012} 
 
\offprints{Martin Groenewegen}

\abstract {
Mass loss is one of the fundamental properties of asymptotic giant branch (AGB) stars, 
but for stars with initial masses below $\sim$ 1 \msol, the mass loss on the first red giant branch (RGB)
actually dominates mass loss on the AGB. Nevertheless, mass loss on the RGB is still often parameterised 
by a simple Reimers law in stellar evolution models.
}
{To study the infrared excess and mass loss of a sample of nearby RGB stars with reliably measured \it Hipparcos \rm parallaxes and compare 
the mass loss to that derived for luminous stars in clusters.}
{The spectral energy distributions  of a well-defined sample of 54 RGB stars are constructed, 
and fitted with the dust radiative transfer model DUSTY. The central stars are modelled by MARCS model atmospheres. 
In a first step, the best-fit MARCS model is derived, basically determining the effective temperature.
In a second step, models with a finite dust optical depth are fitted and it is determined whether the reduction 
in $\chi^2$ in such models  with one additional free parameter is statistically significant.}
{Among the 54 stars, 23 stars are found to have a significant infrared excess, which is interpreted as mass loss.
The most luminous star with $L= 1860\;\lsol$ is found to undergo mass loss, while \rm none of the 5 stars with $L < 262\;\lsol$ display evidence of mass loss. 
In the range 265 $< L < 1500 \;\lsol$, 22 stars out of 48 experience mass loss, which supports the notion of 
episodic mass loss.
It is the first time that excess emission is found in stars fainter than $\sim$600 \lsol.
The dust optical depths are translated into mass-loss rates \it assuming \rm a typical expansion velocity 
of 10 \ks\ and a dust-to-gas ratio of 0.005.
In this case, fits to the stars with an excess result in 
$\log \dot{M}$ (\msolyr) $ = (1.4 \pm 0.4) \log L + (-13.2 \pm 1.2)$ and 
$\log \dot{M}$ (\msolyr) $ = (0.9 \pm 0.3) \log (L \; R/M) + (-13.4 \pm 1.3)$ assuming a mass of 1.1 M$_{\odot}$ for all objects. 
We caution that if the expansion velocity and dust-to-gas ratio have different values from those assumed, 
the constants in the fit will change. If these parameters are also functions of luminosity, then this would 
affect both the slopes and the offsets.
The mass-loss rates are compared to those derived for luminous stars in globular clusters, 
by fitting both the infrared excess, as in the present paper, and the chromospheric lines. 
There is excellent agreement between these values and the mass-loss rates derived from the chromospheric activity.
There is a systematic difference with the literature mass-loss rates derived from modelling the infrared excess, 
and this has been traced to technical details on how the DUSTY radiative transfer model is run.
If the present results are combined with those from modelling the chromospheric emission lines, we obtain the fits
$\log \dot{M}$ (\msolyr) $ = (1.0 \pm 0.3) \log L + (-12.0 \pm 0.9)$ and 
$\log \dot{M}$ (\msolyr) $ = (0.6 \pm 0.2) \log (L \; R/M) + (-11.9 \pm 0.9)$, and find that the metallicity dependence is weak at best.
The predictions of these mass-loss rate formula are tested against the recent RGB mass loss determination in NGC 6791.
Using a scaling factor of $\sim$8 $\pm$ $\sim$5, both relations can fit this value. 
That the scaling factor is larger than unity suggests that the expansion velocity and/or dust-to-gas ratio, or even the dust opacities, 
are different from the values adopted. 
Angular diameters are presented for the sample. They may serve as calibrators in interferometric observations.
}
{}


\maketitle

\section{Introduction} 

Almost all stars with masses between 1 and 8 \msol\ pass through the
asymptotic giant branch (AGB). On the AGB, the mass-loss rate exceeds
the nuclear burning rate, implying that the mass-loss process
dominates stellar evolution. Although not understood in all its
details, the relevant process is believed to be that of
pulsation-enhanced dust-driven winds: shock waves created by stellar
pulsation lead to a dense, cool, extended stellar atmosphere, allowing
for efficient dust formation. The grains are accelerated away from the
star by radiation pressure, dragging the gas along (see the various
contributions in Habing \& Olofsson 2003 for an overview).
Following the advent of Spitzer, an analysis of 200 carbon- and oxygen-rich
AGB stars in the Small and Large Magellanic Clouds with Spitzer IRS
spectra show a clear relation between the mass-loss rate and both the pulsation
period and the luminosity (Groenewegen et al. 2009 and
references therein), confirming earlier work on Galactic stars.

The focus of the present paper however is  mass loss on the first red giant branch (RGB).  
For stars with initial masses of $\la$ 2.2 \msol, this is a prominent evolutionary 
phase where stars reach high luminosities ($\log (L/ L_\odot ) \sim 3$).
Low- and intermediate-mass stars must lose about 0.2 \msol\ on the
RGB in order to explain the morphology on the horizontal branch
(e.g. Catelan 2000 and references therein) and the pulsation
properties of RR Lyrae stars (e.g. Caloi \& d'Antona 2008).  For the
lowest initial masses ($\la$ 1 \msol), the total mass lost on the RGB
dominates that of the AGB phase, and therefore it is equally important
to understand how this process develops. In stellar evolutionary
models, the RGB mass loss is often parameterised by the Reimers law
(1975) with some scaling parameter (typically $\eta \sim 0.4$).

The RGB mass loss can arise from chromospheric activity (see Mauas et al. 2006, 
M\'esz\'aros et al. 2009, Vieytes et al. 2011) or can also be pulsation-enhanced and dust-driven. 
The studies of Boyer et al. (2010), Origlia et al. (2007, 2010), and
Momany et al. (2012) of 47 Tuc, and McDonald et al. (2009, 2011) of
$\omega$ Cen illustrate the current state of affairs regarding the
dust modelling. Boyer et al. (2010) uses Spitzer 3.6 and 8 $\mu$m data
to show that the reddest colours on the RGB are reached at the tip of
the RGB (TRGB), and that these are known long period variables
(Clement et al. 2001, Lebzelter \& Wood 2005) with periods between 50
and 220 days.  In McDonald et al. (2009), optical and NIR data is
combined with Spitzer IRAC and MIPS 24 $\mu$m data to model the
spectral energy distributions (SEDs) and derive mass-loss rates. They
conclude that two-thirds of the total mass loss is by dusty winds and
one-third by chromospheric activity. They show that the highest (dust)
mass-loss rates occur near the TRGB, which, indeed, are known variable
stars being mostly semi-regular pulsators.  At lower luminosities
along the RGB, there is an excess at 24 $\mu$m that can be translated
into a mass-loss rate but the uncertainties are large.
In an alternative approach, using asteroseismology to estimate the
mass of red clump stars, and stars on the RGB fainter than the luminosity of
the clump, Miglio et al. (2012) estimate the amount of mass lost in
NGC 6791 and NGC 6819, and conclude that it is consistent with a Reimers law with 0.1 $\la \eta \la$ 0.3.

In the present paper, a complimentary approach is taken by studying the infra-red excess around 
nearby RGB stars, based on a sample of stars with accurate parallaxes.
In Sect.~2, we present the sample in addition to the photometric data used to constrain the modelling.
In Sect.~3, the dust radiative transfer models are introduced, and the results are presented in Sect.~4.
Our results are discussed in Sect.~5.

\section{The sample} 

We selected our sample from the Hipparcos catalog. The
parallaxes were taken from van Leeuwen (2007), and other data for
the stars was gathered from the original release (ESA 1997).
In a first step, supposedly single stars were selected where the parameter fit was good
(Hipparcos flags isoln = 5 and gof$<$ 5.0).
To ensure an accurate determination of the luminosity, a positive
parallax and relative error of smaller than 10\% were required
($\pi > 0$, $\sigma_{\pi}/ \pi < 0.1$).

\begin{figure} 
\resizebox{\hsize}{!}{\includegraphics{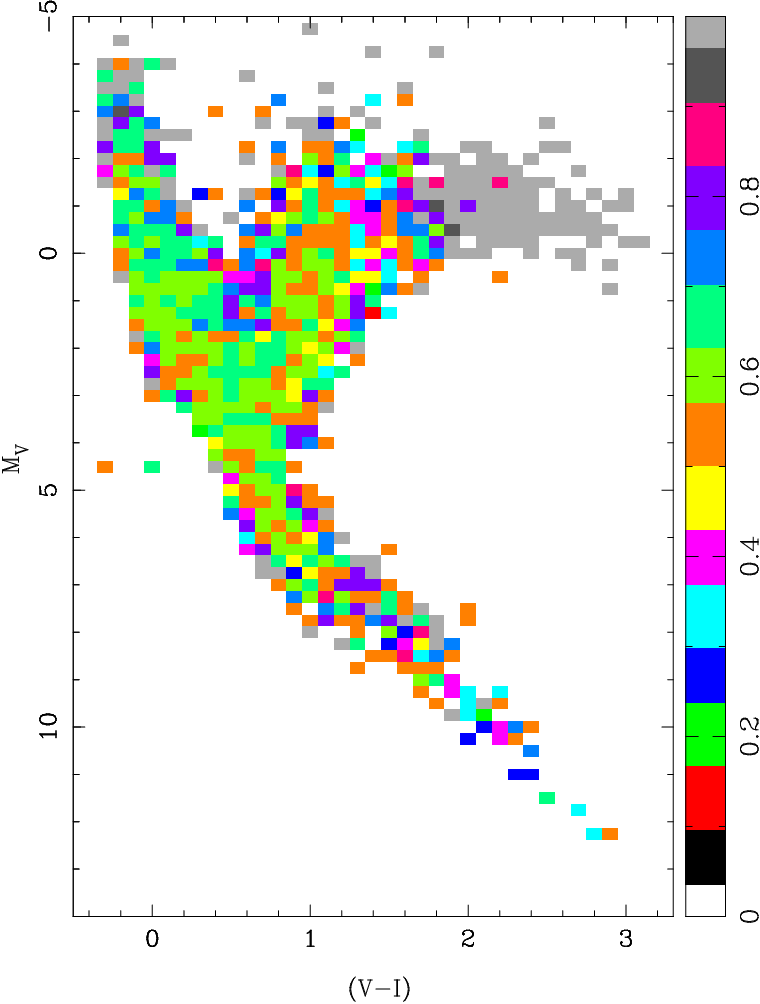}}
\caption[]{ 
The fraction of variable stars 
in the Hipparcos catalog based on the van Leeuwen (2007) parallax data.
The bin size is 0.1 in $(V-I)$ and 0.25 in $M_{\rm V}$. 
Considered are stars with an relative parallax error $<0.15$ and an error in  $(V-I) < 0.15$ magnitude.
The RGB stands out as having almost 100\% variability.
} 
\label{fig-hipvar} 
\end{figure}

It is well-established (see e.g. McDonald et al. 2009) that mass loss is
larger in stars near the tip of the RGB but also that these stars often show variability.
To illustrate this, Figure~\ref{fig-hipvar} shows the fraction of variable stars across the Hertzsprung-Russell diagram.
We plot 44177 stars from Hipparcos that have a parallax error of smaller than 15\% and an error in $(V-I)$ of smaller than 0.15 mag.
The cells have a width of 0.1 magnitude in $(V-I)$ and 0.25 magnitude in $M_{\rm V}$. 
The RGB is clearly visible with a very high fraction of variables.
On the basis of this, a further selection of $(V-I) > 1.5$ (and $\sigma_{\rm (V-I)} < 0.1$) was imposed.
The possible effect of using this lower limit to the $(V-I)$ colour on the fraction of mass-losing RGB stars is discussed in Sect.~5.1.

At this point in the selection, the interstellar reddening $A_{\rm V}$
is determined using the method outlined in Groenewegen (2008), which
combines various three-dimensional estimates of $A_{\rm V}$ (Marshall et al. 2006, Drimmel et al. 2003, and Arenou et al. 1992).

Following Koen \& Laney (2000, see also Dumm \& Schild 1998), the effective temperature is estimated from the relation
$\log T_{\rm eff}$ (K)$ = 3.65 - 0.035 \cdot {(V-I)_{\rm 0}}$
and the stellar radius (in solar units) from
$\log R = 2.97 - \log (\pi) -0.2 \cdot (V_{\rm o} + 0.356 \cdot {(V-I)_0})$,
from which the luminosity in solar units is then determined.
A final selection using
$M_{\rm V} < +1.0$ (see Figure~\ref{fig-hipvar}), 100 $< L <$ 2000 \lsol, and $A_{\rm V} < 0.1$ is
applied, again to ensure the selection of giants and that reddening
will play no role in the interpretation of the results.

In the section below the methodology is outlined but  is based on
fitting models to the spectral energy distribution (SED).  Therefore, the availability of sufficient photometric data is crucial, especially
in the optical and near-infrared as this will be the main constraint
in deriving the best-fit model atmosphere.
For this reason, stars that lacked either optical $UBV$ and/or $JHK$ data were removed from the sample.
In addition, stars that have an IRAS CIRR3 flag $>$25 MJy/sr were also removed, in order 
to avoid contamination by cirrus of the IRAS 60 and 100 $\mu$m data.

The sample thus selected contains 54 objects (15 of spectral type K, and 39 M giants) 
of which the basic properties have been listed in Table~\ref{Tab-sample}.

Most data listed come from the Hipparcos catalog, including the type of variability and
the difference between the 95\% and 5\% percentile values of $H_{\rm P}$ ($\Delta H_{\rm P}$).
In addition, the reddening (see above, col.~4), the spectral type (from SIMBAD, col.~7), and  
both [Fe/H] and $\log g$ determinations from the literature (cols.~10-11) are listed.

\begin{table*}


\caption{The RGB sample.}

  \begin{tabular}{rrrrrrrrrrrr}
  \hline \hline
HIP &   parallax  &  $M_{\rm V}$ &  $A_{\rm V}$  & $V$  & $(V-I)$ & spectral type  & variability  & $\Delta H_{\rm P}$ & [Fe/H] &  $\log g^b$ \\
    &   (mas)     &             &              &      &         &                & type$^a$     & (mag)             &        &  \\ 
\hline

 4147 & 5.56 $\pm$ 0.22 & $-$1.58 & 0.09 & 4.78 & 1.66 & M0III & M & 0.03 &  \\
12107 & 7.00 $\pm$ 0.42 & $-$0.33 & 0.08 & 5.53 & 1.98 & M0III & U & 0.05 &  \\
32173 & 10.08 $\pm$ 0.33 & $-$0.01 & 0.06 & 5.04 & 1.50 & K5III & C & 0.02 & $-0.03$ & 1.5 (1) \\ 
37300 & 6.21 $\pm$ 0.29 & $-$1.07 & 0.07 & 5.04 & 1.72 & M0III & U & 0.03 &  \\
37946 & 7.51 $\pm$ 0.41 & $-$0.55 & 0.08 & 5.15 & 2.03 & M3III & U & 0.04 &  \\
41822 & 7.88 $\pm$ 0.30 & $-$0.24 & 0.05 & 5.33 & 1.57 & K5III & U & 0.03 &  \\
44126 & 5.33 $\pm$ 0.44 & $-$0.16 & 0.09 & 6.30 & 2.48 & M4III & U & 0.10 &  \\
44390 & 10.36 $\pm$ 0.25 & $-$0.23 & 0.04 & 4.74 & 2.15 & M3III & U & 0.04 & \\
44857 & 6.25 $\pm$ 0.30 & $-$0.93 & 0.06 & 5.15 & 1.55 & K5III & M & 0.03 &  $-0.23$ & 1.66 (2) \\ 
46750 & 9.92 $\pm$ 0.18 & $-$0.74 & 0.04 & 4.32 & 1.63 & K5III & M & 0.02 &  $-0.29$ & 1.6 (2) \\ 
47723 & 5.35 $\pm$ 0.33 & $-$1.08 & 0.08 & 5.36 & 1.94 & M2III & U & 0.06 &  \\
49005 & 6.51 $\pm$ 0.35 & $-$0.49 & 0.06 & 5.50 & 1.51 & K5III & M & 0.02 &  \\
49029 & 8.04 $\pm$ 0.29 & $-$0.85 & 0.05 & 4.68 & 1.96 & M2III & U & 0.03 &  \\
52366 & 4.02 $\pm$ 0.33 & $-$1.04 & 0.08 & 6.02 & 2.45 & M2III & U & 0.09 &  \\
52863 & 5.99 $\pm$ 0.42 & $-$0.26 & 0.06 & 5.92 & 1.91 & M2III & M & 0.04 &  \\
53449 & 8.42 $\pm$ 0.37 & +0.49 & 0.05 & 5.91 & 3.50 & M5.5I & P & 0.26 &  & 2.3 (1) \\ 
53726 & 3.96 $\pm$ 0.38 & $-$1.10 & 0.08 & 5.99 & 2.17 & M2III & U & 0.06 &  \\
53907 & 5.57 $\pm$ 0.24 & $-$1.61 & 0.07 & 4.73 & 1.77 & M0III & U & 0.03 & $-0.23$ & 1.22 (3)  \\ 
54537 & 5.96 $\pm$ 0.50 & $-$0.31 & 0.08 & 5.89 & 2.16 & M2III & U & 0.06 &  \\
55687 & 8.67 $\pm$ 0.22 & $-$0.55 & 0.05 & 4.81 & 1.67 & K5III & M & 0.03 & $-0.38$ & 1.61 (2) \\ 
56127 & 5.38 $\pm$ 0.31 & $-$1.64 & 0.07 & 4.77 & 1.62 & K3.5I & M & 0.03 & $-0.31$ & 1.80 (3) \\ 
56211 & 9.80 $\pm$ 0.16 & $-$1.27 & 0.04 & 3.82 & 1.79 & M0III & U & 0.05 &  \\
60122 & 5.51 $\pm$ 0.28 & $-$1.08 & 0.07 & 5.28 & 1.90 & M0III & U & 0.04 &  \\
60795 & 7.82 $\pm$ 0.32 & +0.09 & 0.06 & 5.68 & 1.88 & M2III & M & 0.03 &  \\
61658 & 6.64 $\pm$ 0.31 & $-$0.29 & 0.08 & 5.68 & 2.32 & M3III & U & 0.07 &  \\
62443 & 5.59 $\pm$ 0.45 & +0.08 & 0.08 & 6.42 & 2.14 & M4III & U & 0.05 &  \\
63355 & 10.06 $\pm$ 0.28 & $-$0.28 & 0.06 & 4.76 & 1.79 & M1III & M & 0.03 &   & 1.0 (1) \\ 
64607 & 6.82 $\pm$ 0.32 & $-$0.26 & 0.07 & 5.64 & 1.53 & M0III & U & 0.05 &  \\
66417 & 7.12 $\pm$ 0.34 & $-$0.09 & 0.07 & 5.72 & 1.96 & M2III & U & 0.05 &  \\
66738 & 6.23 $\pm$ 0.22 & $-$1.47 & 0.07 & 4.63 & 1.97 & M2III & U & 0.06 &  $+0.30$ & 1.60 (4) \\ 
67605 & 4.80 $\pm$ 0.38 & $-$0.79 & 0.09 & 5.89 & 1.94 & M2III & U & 0.05 &  \\
67627 & 8.79 $\pm$ 0.20 & $-$0.75 & 0.05 & 4.58 & 2.35 & M3.5I & U & 0.09 & $-0.24$ & 1.25 (3) \\ 
67665 & 5.43 $\pm$ 0.20 & $-$1.65 & 0.08 & 4.76 & 1.63 & K5III & U & 0.12 & $+0.50$ & 0.50 (2)\\ 
69068 & 5.95 $\pm$ 0.25 & $-$0.94 & 0.08 & 5.26 & 1.97 & M1.5I & U & 0.08 &  \\
69373 & 7.62 $\pm$ 0.19 & $-$0.46 & 0.05 & 5.18 & 1.93 & M2III & U & 0.03 &  \\
71280 & 3.82 $\pm$ 0.26 & $-$1.44 & 0.09 & 5.74 & 1.71 & M1III & U & 0.04 & \\
73568 & 8.78 $\pm$ 0.28 & $-$0.55 & 0.07 & 4.80 & 1.54 & K4III & M & 0.02 &  $-0.04$ & 1.68 (2)\\ 
76307 & 5.26 $\pm$ 0.24 & $-$1.35 & 0.10 & 5.14 & 1.90 & M1.5I & U & 0.05 &  \\
77661 & 8.72 $\pm$ 0.30 & $-$0.62 & 0.06 & 4.74 & 1.60 & K5III & M & 0.03 & $-0.17$ & 1.68 (2) \\ 
78632 & 3.40 $\pm$ 0.33 & $-$1.24 & 0.09 & 6.19 & 1.87 & M1III & U & 0.04 &  \\
80042 & 4.32 $\pm$ 0.43 & $-$0.34 & 0.09 & 6.57 & 1.57 & M2III & U & 0.06 & \\
80197 & 5.08 $\pm$ 0.22 & $-$1.37 & 0.10 & 5.20 & 2.00 & M2III & U & 0.05 &  \\
80214 & 5.48 $\pm$ 0.24 & $-$1.00 & 0.09 & 5.40 & 1.57 & K5III & M & 0.03 &  $-0.16$ & 1.76 (2) \\ 
82073 & 9.21 $\pm$ 0.41 & $-$0.10 & 0.07 & 5.15 & 1.54 & K5III & M & 0.03 &  $-0.03$ & 1.52 (3) \\ 
83430 & 8.26 $\pm$ 0.26 & $-$0.52 & 0.07 & 4.97 & 2.08 & M3III & U & 0.05 &  \\
84835 & 5.84 $\pm$ 0.19 & $-$0.75 & 0.10 & 5.51 & 1.71 & M0III & M & 0.04 &  \\
87833 & 21.15 $\pm$ 0.10 & $-$1.17 & 0.03 & 2.24 & 1.54 & K5III & M & 0.02 & $-0.05$ & 1.32 (3) \\ 
88122 & 5.71 $\pm$ 0.25 & $-$0.63 & 0.10 & 5.69 & 1.69 & M0III & M & 0.03 &  \\
98401 & 6.20 $\pm$ 0.31 & +0.08 & 0.08 & 6.20 & 2.09 & M3III & U & 0.06 &  \\
106140 & 8.29 $\pm$ 0.19 & $-$0.97 & 0.08 & 4.52 & 1.82 & M1III & U & 0.04 &  \\
112716 & 10.28 $\pm$ 0.29 & $-$0.97 & 0.08 & 4.05 & 1.72 & K5III & M & 0.03 &  \\
114144 & 9.92 $\pm$ 0.29 & $-$0.54 & 0.06 & 4.54 & 1.79 & M1III & U & 0.04 &  & 1.0 (1) \\ 
115669 & 11.50 $\pm$ 0.22 & $-$0.38 & 0.06 & 4.38 & 1.52 & K4III & M & 0.02 &  $-0.20$ & 1.66 (2) \\ 
117718 & 7.07 $\pm$ 0.27 & $-$0.79 & 0.10 & 5.06 & 2.09 & M2III & U & 0.06 &  \\

\hline
\end{tabular}

{\bf Notes.}
$^{(a)}$ Variability type, Field H52 in the Hipparcos (ESA 1997) catalog. Meaning:
C = 'constant', not detected as being variable, 
D = duplicity-induced variability flag, 
M = possible micro variable, 
U = unsolved variable, 
P = Periodic, 
R = revised colour index.

$^{(b)}$ References for [Fe/H] and $\log g$: 
(1) Massarotti et al. (2008) and references therein; 
(2) McWilliam (1990);
(3) Cenarro et al. (2007);
(4) Fern\'andez-Villaca\~nas et al. (1990).

\label{Tab-sample}
\end{table*}

\begin{figure*}[!ht]

\begin{minipage}{0.49\textwidth}
\resizebox{\hsize}{!}{\includegraphics{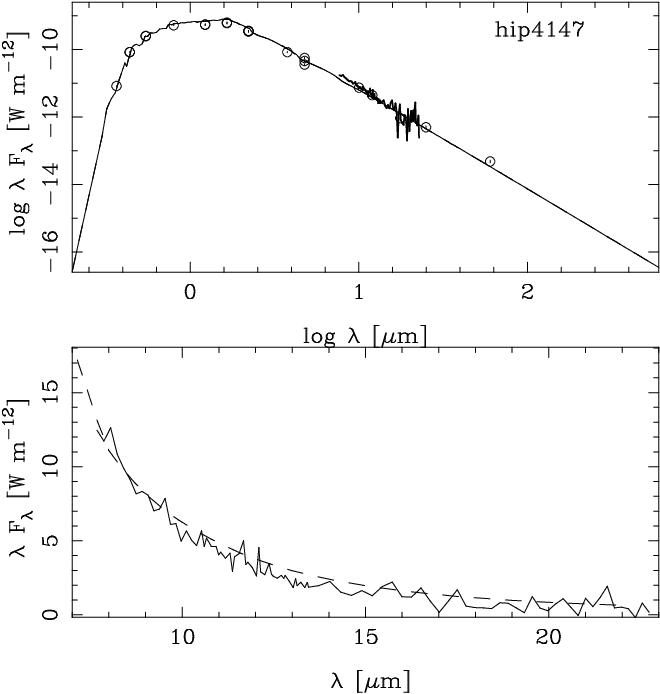}}
\end{minipage}
\begin{minipage}{0.49\textwidth}
\resizebox{\hsize}{!}{\includegraphics{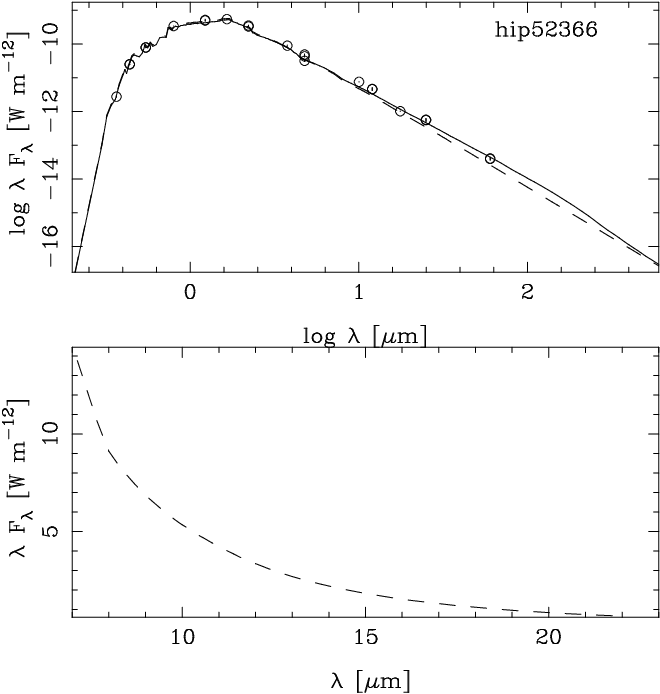}}
\end{minipage}

\caption[]{ 
Example fits to the SED (top panel) and IRAS LRS spectra (lower panel).
In the top panel, the solid line indicates the best fit, the dashed line the model without mass loss.
For HIP 4147, the best-fit model {\it is} the model without mass loss so that the two lines are over-plotted.
The observed photometry is plotted by the circles, and error bars are also plotted, but are typically much smaller than the symbol size.
In the lower panel, the best-fit model is indicated by the dashed line, and the LRS spectrum by the solid line. 
Sometimes no LRS spectrum is available, as for HIP 52366.
The complete figure is available in Appendix~A in the electronic edition.
} 
\label{Fig-SED} 
\end{figure*}

\begin{table*}


\caption{The RGB sample: fit results}

\centering
  \begin{tabular}{rrrrrlrrrr}
  \hline \hline
HIP &   $T_{\rm eff}$ &  Dust           &            $L$    &         $\theta$     & $\tau_{\rm V}$ & \mdot      & code & $\chi^2_{\rm r}$ \\
    &    (K)         &                 &           (\lsol) &          (mas)       &               & (\msolyr) &      &       \\ 
\hline

  4147  & 3800 &                    &    1044  $\pm$ 25  &  3.85  $\pm$  0.15  & 10$^{-5}$  & 1.05 \, 10$^{-12}$ & 0 &      74  \\
 12107  & 3700 &                    &     482  $\pm$ 18  &  3.47  $\pm$  0.15  & 10$^{-5}$  & 4.88 \, 10$^{-13}$ & 0 &     278  \\
 32173  & 3900 &                    &     219  $\pm$ 09  &  3.03  $\pm$  0.13  & 10$^{-5}$  & 4.95 \, 10$^{-13}$ & 0 &     228  \\
 37300  & 3800 &                    &     665  $\pm$ 10  &  3.43  $\pm$  0.13  & 10$^{-5}$  & 8.41 \, 10$^{-13}$ & 0 &      13  \\
 37946  & 3700 &                    &     599  $\pm$ 15  &  4.15  $\pm$  0.17  & 10$^{-5}$  & 5.44 \, 10$^{-13}$ & 0 &      70  \\
 41822  & 3900 &                    &     259  $\pm$ 11  &  2.58  $\pm$  0.11  & 10$^{-5}$  & 5.38 \, 10$^{-13}$ & 0 &     189  \\
 44126  & 3600 &          iron      &     528  $\pm$ 16  &  2.92  $\pm$  0.12  & (4.99 $\pm$ 0.15) \, 10$^{-2}$ & 2.54 \, 10$^{-09}$ & 1 &      46  \\
 44390  & 3600 &          iron      &     522  $\pm$ 14  &  5.65  $\pm$  0.23  & (8.36 $\pm$ 0.94) \, 10$^{-3}$ & 4.21 \, 10$^{-10}$ & 1 &      34  \\
 44857  & 3800 &                    &     560  $\pm$ 20  &  3.17  $\pm$  0.13  & 10$^{-5}$  & 7.72 \, 10$^{-13}$ & 0 &     202  \\
 46750  & 3800 &                    &     467  $\pm$ 10  &  4.59  $\pm$  0.18  & 10$^{-5}$  & 7.05 \, 10$^{-13}$ & 0 &      60  \\
 47723  & 3700 &                    &     842  $\pm$ 24  &  3.51  $\pm$  0.14  & 10$^{-5}$  & 9.21 \, 10$^{-13}$ & 0 &      97  \\
 49005  & 4000 &                    &     293  $\pm$ 04  &  2.15  $\pm$  0.08  & 10$^{-5}$  & 5.85 \, 10$^{-13}$ & 0 &       8  \\
 49029  & 3700 &          iron      &     688  $\pm$ 18  &  4.76  $\pm$  0.19  & (8.04 $\pm$ 0.74) \, 10$^{-3}$ & 4.69 \, 10$^{-10}$ & 1 &      31  \\
 52366  & 3600 &          iron      &    1207  $\pm$ 32  &  3.33  $\pm$  0.14  & (3.56 $\pm$ 0.15) \, 10$^{-2}$ & 2.73 \, 10$^{-09}$ & 1 &      28  \\
 52863  & 3700 &                    &     384  $\pm$ 13  &  2.65  $\pm$  0.11  & 10$^{-5}$  & 6.22 \, 10$^{-13}$ & 0 &      82  \\
 53449  & 3300 &          iron      &    1346  $\pm$ 44  &  8.77  $\pm$  0.40  & (5.58 $\pm$ 0.20) \, 10$^{-2}$ & 4.41 \, 10$^{-09}$ & 1 &      41  \\
 53726  & 3600 &                    &    1071  $\pm$ 32  &  3.09  $\pm$  0.13  & 10$^{-5}$  & 7.20 \, 10$^{-13}$ & 0 &      99  \\
 53907  & 3800 &                    &    1095  $\pm$ 34  &  3.95  $\pm$  0.16  & 10$^{-5}$  & 1.08 \, 10$^{-12}$ & 0 &     107  \\
 54537  & 3700 &                    &     410  $\pm$ 16  &  2.73  $\pm$  0.12  & 10$^{-5}$  & 6.43 \, 10$^{-13}$ & 0 &     178  \\
 55687  & 3900 &          iron      &     353  $\pm$ 08  &  3.31  $\pm$  0.12  & (5.62 $\pm$ 5.16) \, 10$^{-4}$ & 2.39 \, 10$^{-11}$ & 1 &      30  \\
 56127  & 3900 &          iron      &     975  $\pm$ 55  &  3.41  $\pm$  0.16  & (8.08 $\pm$ 0.51) \, 10$^{-3}$ & 5.72 \, 10$^{-10}$ & 1 &     177  \\
 56211  & 3700 &                    &     959  $\pm$ 16  &  6.85  $\pm$  0.27  & 10$^{-5}$  & 9.83 \, 10$^{-13}$ & 0 &      32  \\
 60122  & 3700 &                    &     836  $\pm$ 20  &  3.60  $\pm$  0.14  & 10$^{-5}$  & 6.43 \, 10$^{-13}$ & 0 &      61  \\
 60795  & 3800 &                    &     246  $\pm$ 06  &  2.62  $\pm$  0.10  & 10$^{-5}$  & 5.12 \, 10$^{-13}$ & 0 &      69  \\
 61658  & 3600 &          iron      &     601  $\pm$ 22  &  3.88  $\pm$  0.17  & (1.94 $\pm$ 0.088) \, 10$^{-2}$  & 1.05 \, 10$^{-09}$ & 1 &      70  \\
 62443  & 3700 &                    &     279  $\pm$ 14  &  2.11  $\pm$  0.10  & 10$^{-5}$  & 5.31 \, 10$^{-13}$ & 0 &     363  \\
 63355  & 3800 &          iron      &     324  $\pm$ 12  &  3.88  $\pm$  0.16  & (4.14 $\pm$ 0.70) \, 10$^{-3}$  & 1.67 \, 10$^{-10}$ & 1 &      50  \\
 64607  & 3900 &          iron      &     267  $\pm$ 13  &  2.26  $\pm$  0.10  & (1.28 $\pm$ 0.061) \, 10$^{-2}$ & 4.75 \, 10$^{-10}$ & 1 &     129  \\
 66417  & 3700 &                    &     336  $\pm$ 07  &  2.95  $\pm$  0.12  & 10$^{-5}$  & 5.82 \, 10$^{-13}$ & 0 &      46  \\
 66738  & 3600 &                    &    1458  $\pm$ 32  &  5.67  $\pm$  0.23  & 10$^{-5}$  & 8.40 \, 10$^{-13}$ & 0 &      55  \\
 67605  & 3800 &          iron      &     545  $\pm$ 13  &  2.40  $\pm$  0.09  & (1.47 $\pm$ 0.13) \, 10$^{-2}$  & 7.72 \, 10$^{-10}$ & 1 &      26  \\
 67627  & 3500 &                    &    1113  $\pm$ 19  &  7.40  $\pm$  0.30  & 10$^{-5}$  & 7.26 \, 10$^{-13}$ & 0 &      31  \\
 67665  & 3600 &          iron      &    1895  $\pm$ 73  &  5.64  $\pm$  0.24  & (3.96 $\pm$ 0.096) \, 10$^{-2}$ & 3.81 \, 10$^{-09}$ & 1 &      61  \\
 69068  & 3700 &          iron      &     790  $\pm$ 23  &  3.78  $\pm$  0.15  & (1.66 $\pm$ 0.059) \, 10$^{-2}$  & 1.04 \, 10$^{-09}$ & 1 &      61  \\
 69373  & 3700 &          iron      &     477  $\pm$ 17  &  3.76  $\pm$  0.16  & (5.18 $\pm$ 0.63) \, 10$^{-3}$& 2.52 \, 10$^{-10}$ & 1 &      68  \\
 71280  & 3900 &          iron      &     818  $\pm$ 76  &  2.22  $\pm$  0.13  & (1.45 $\pm$ 0.065) \, 10$^{-2}$  & 9.42 \, 10$^{-10}$ & 1 &     426  \\
 73568  & 4000 &                    &     318  $\pm$ 09  &  3.02  $\pm$  0.11  & 10$^{-5}$  & 6.09 \, 10$^{-13}$ & 0 &     101  \\
 76307  & 3800 &          iron      &     937  $\pm$ 36  &  3.45  $\pm$  0.14  & (2.60 $\pm$ 0.71) \, 10$^{-3}$  & 1.79 \, 10$^{-10}$ & 1 &      80  \\
 77661  & 3900 &       AlOx         &     358  $\pm$ 06  &  3.35  $\pm$  0.12  & (5.00 $\pm$ 1.59) \, 10$^{-4}$  & 3.16 \, 10$^{-11}$ & 1 &      24  \\
 78632  & 3800 &                    &     844  $\pm$ 18  &  2.11  $\pm$  0.08  & 10$^{-5}$  & 9.45 \, 10$^{-13}$ & 0 &      29  \\
 80042  & 3900 &          iron      &     302  $\pm$ 07  &  1.52  $\pm$  0.06  & (6.69 $\pm$ 0.17) \, 10$^{-2}$ & 2.66 \, 10$^{-09}$ & 1 &      30  \\
 80197  & 3700 &          iron      &    1157  $\pm$ 32  &  3.90  $\pm$  0.16  & (1.50 $\pm$ 0.064) \, 10$^{-2}$ & 1.14 \, 10$^{-09}$ & 1 &      47  \\
 80214  & 4000 &                    &     426  $\pm$ 22  &  2.19  $\pm$  0.10  & 10$^{-5}$  & 7.06 \, 10$^{-13}$ & 0 &     305  \\
 82073  & 3900 &                    &     224  $\pm$ 05  &  2.80  $\pm$  0.11  & 10$^{-5}$  & 5.01 \, 10$^{-13}$ & 0 &      63  \\
 83430  & 3700 &          iron      &     526  $\pm$ 10  &  4.28  $\pm$  0.17  & (1.71 $\pm$ 0.11) \, 10$^{-2}$  & 8.74 \, 10$^{-10}$ & 1 &      23  \\
 84835  & 3800 &                    &     455  $\pm$ 15  &  2.67  $\pm$  0.11  & 10$^{-5}$  & 6.96 \, 10$^{-13}$ & 0 &     157  \\
 87833  & 3900 &                    &     602  $\pm$ 04  & 10.55  $\pm$  0.38  & 10$^{-5}$  & 8.20 \, 10$^{-13}$ & 0 &       4  \\
 88122  & 4000 &          iron      &     351  $\pm$ 15  &  2.07  $\pm$  0.08  & (4.11 $\pm$ 0.084) \, 10$^{-2}$ & 1.77 \, 10$^{-09}$ & 1 &      60  \\
 98401  & 3700 &          iron      &     285  $\pm$ 13  &  2.37  $\pm$  0.10  & (5.59 $\pm$ 0.67) \, 10$^{-3}$ & 2.10 \, 10$^{-10}$ & 1 &     121  \\
106140  & 3700 &                    &     682  $\pm$ 12  &  4.89  $\pm$  0.19  & 10$^{-5}$  & 8.29 \, 10$^{-13}$ & 0 &      33  \\
112716  & 3900 &                    &     567  $\pm$ 13  &  4.98  $\pm$  0.19  & 10$^{-5}$  & 7.96 \, 10$^{-13}$ & 0 &      54  \\
114144  & 3900 &          iron      &     373  $\pm$ 08  &  3.90  $\pm$  0.15  & (7.07 $\pm$ 8.43) \, 10$^{-4}$ & 3.10 \, 10$^{-11}$ & 1 &      34  \\
115669  & 4000 &                    &     262  $\pm$ 04  &  3.60  $\pm$  0.13  & 10$^{-5}$  & 3.70 \, 10$^{-13}$ & 0 &      10  \\
117718  & 3600 &                    &     816  $\pm$ 19  &  4.82  $\pm$  0.20  & 10$^{-5}$  & 6.29 \, 10$^{-13}$ & 0 &      68  \\

\hline
\end{tabular}

\tablefoot{
Listed are,
(Col~2) the effective temperature of the MARCS model, 
(Col~3) the dust component that fits best in the case when $\tau_{\rm V} \not= 10^{-5}$, 
(Col~4) luminosity and the internal error (that is, the error in the Hipparcos distance is not included), 
(Col~5) the angular diameter in mas and the error (based on the error in $L$, and a 70~K error in $T_{\rm eff}$), 
(Col~6) the dust optical depth in the $V$-band and the internal error, and 
(Col~7) the corresponding mass-loss rate assuming a constant expansion velocity of 10 \ks, 
a dust-to-gas ($\Psi$) ratio of 0.005 and 
grain specific density, $\rho$, of 5.1 g cm$^{-3}$ (appropriate for iron grains in a DHS with 70\% vacuum), 
(Col~8) a code if the optical depth was fitted (1) or fixed (0), 
(Col~9) the reduced $\chi^2$ to indicate the goodness of the fit.
}

\label{Tab-fit}
\end{table*}

\section{The model}
\label{themodel}

The code used in this paper is based on that presented by Groenewegen et al. (2009). 
In that paper, a dust radiative transfer model was included as a subroutine in a
minimisation code using the the {\sc mrqmin} routine (using the
Levenberg-Marquardt method from Press et al.\ 1992). The parameters that
were fitted in the minimisation process include the dust optical depth in the $V$-band ($\tau_{\rm V}$), 
luminosity, and the dust temperature at the inner radius ($T_{\rm c}$). 
The output of a model is an SED, which is folded with the relevant
filter response curves to produce magnitudes that can be compared to the
observations (see Groenewegen 2006).
Spectra can also be included in the minimisation process.
In Groenewegen et al. (2009) the dust radiative transfer model was that
of Groenewegen (1993; also see Groenewegen 1995), but this has since been
replaced by the dust radiative transfer model DUSTY (Ivezi\'c et al. 1999).

The central star was modelled by a MARCS stellar photosphere model\footnote{http://marcs.astro.uu.se/} (Gustafsson et al. 2008).  
Models for temperatures between 3200~K and 4000~K 
(in steps of 100~K), and those of 4250~K and 4500~K, with solar metallicity and $\log g$= 1.5 were considered. 
Spectroscopically determined metallicites and gravities were only available for a third of the sample 
(Table~\ref{Tab-sample}) but indicate that $\log g$ = 1.5 is an appropriate value. 
The median metallicity is $-0.17$ dex, so slightly subsolar.

As the mass-loss rates in RGB stars are expected to be small, and the
IRAS LRS spectra (see below) show no hint of a 9.8 $\mu$m silicate
feature, two types of dust were considered: aluminium oxide (AlOx), and metallic iron.
The first species is expected as one of the first condensates in an oxygen-rich environment 
(see Niyogi et al. 2011 for a recent discussion), while metallic iron has gained interest in
the past few years as a source of opacity (see e.g. McDonald et al. 2010).

Absorption and scattering coefficients were calculated for grains of
radius 0.15 $\mu$m in the approximation of a "distribution of hollow spheres" (DHS) 
(Min et al. 2005, a vacuum fraction of 70\% is adopted) using the optical constants of
Begemann et al. (1997) for AlOx, and
Pollack et al. (1994) for iron.

The SEDs were constructed considering the following sources of photometry:
Mermilliod (1991) for $UBV$ photometry, and the $I$ magnitude as listed in the Hipparcos catalog,
Gezari et al. (1999) for $JHKLM$ photometry 
(2MASS was not considered, as owing to the brightness of the sources, 
the 2MASS photometry was either saturated or had very large error bars), 
the IRAS {\it Point Source Catalog} (PSC, Beichman et al. 1985) and 
{\it Faint Source Catalog} (FSC, Moshir et al. 1989) for 12, 25, 60, and 100 $\mu$m data 
(only data with flux-quality 3 were considered), 
{\it Akari} IRC (Ishihara et al. 2010) and FIS (Yamamura et al. 2010) mid- and far-IR data,
In addition, the IRAS LRS spectra (Olnon et al. 1986) available from
Volk \& Cohen (1989)\footnote{http://www.iras.ucalgary.ca/$^\sim$volk/getlrs\_plot.html}
were used when available.  The spectra were typically scaled by
factors 1.3-1.6 to ensure that they agreed with the IRAS 12 $\mu$m and/or Akari S9W filter.

In a first iteration, models with effectively no mass loss ($\tau_{\rm V} = 10^{-5}$) 
were run by varying only the effective temperature. A $r^{-2}$ density distribution was assumed, 
and the condensation temperature was fixed at 1000~K in all models, 
as this cannot be constrained from the current photometric datasets
for low mass-loss rates.  The best-fit model was determined.
In a second iteration, for that effective temperature, models with
fixed optical depths of $\tau_{\rm V} = 10^{-4}, 10^{-3}, 10^{-2}$ were run for both iron and AlOx dust. 
The best-fit model was determined (thus fixing the dust component), and 
then models where $\tau$ was also allowed to vary were run.
Finally, models were run in which $\tau$ was allowed to vary using effective temperatures 
one step cooler and hotter in the available grid of MARCS models.

\section{Results} 
\label{S-Results}

Table~\ref{Tab-fit} lists the parameters of the models that provide the best fit to the observed data. 
The fit error in the derived optical depth is typically small, in the median only 5\%, but in some cases much larger.
However, this error does not take into account e.g. the effect of varying the model atmosphere.
Some tests were performed by varying the gravity of the model atmosphere by $\pm0.5$ dex and the metallicity by $\pm0.25$ dex 
and refitting the optical depth. The results suggest that this represents an additional 50\% uncertainty.
The largest uncertainty is in the conversion from optical depth to mass-loss rate. On the one hand, a systematic error as 
the mean velocity and mean dust-to-gas ratio may differ from the adopted values, and, on the other hand, the values 
for individual stars will scatter around these mean values. 
A random error of a factor of two is adopted in the latter case, and this error dominates the error budget.

Examples of the fits are shown in Fig.~2, and all fits are displayed in Appendix~A.
The panels with the SEDs show the best fit (solid line), and a model without mass loss (dashed line) for comparison.
The differences are often small, certainly visually, but are statistically significant.

From inspecting the plots, it is also clear that far-IR data ($\sim$ 80-200 $\mu$m) 
would certainly be very valuable in constraining the models,
as any excess is expected to be largest in that wavelength region.  
In this respect, it is unfortunate that the \it Akari/FIS \rm has
relatively poor sensitivity.  All 54 stars have \it Akari/IRC \rm S9W
and L18W data, but only 15 have FIS WS-band data at 90 $\mu$m and none
of the stars are detected in the filters at 140 and 160 $\mu$m.
None of the stars are detected in the \it Planck \rm Early Release Compact Source Catalogue (The Planck collaboration 2011), 
no appear to have MIPS or Herschel data.

In addition, high-quality mid-IR spectra would also be useful to
improve upon the, in most cases, relatively poor quality IRAS LRS spectrum. 
Only for one object does an ISO SWS spectrum exist (HIP 87833 = $\gamma$ Dra). 
With the current data, no clear dust feature is visible in any of the stars, hence, when a significant infrared excess detected,
the best fit is provided by the featureless metallic iron model rather than the aluminium oxide one (except in one case).

Other mechanisms can produce an infrared excess that is not due to dust emission, as discussed in McDonald et al. (2010), 
e.g.  free-free emission or emission from shells of molecular gas (a MOLsphere; e.g. Tsuji 2000).  
Even featureless dust emission could in principle also be due to
extremely large ($\sim$50 $\mu$m) silicate or amorphous carbon grains 
(McDonald et al. 2010), but these species are not really expected to condense and
form first in these low-density oxygen-rich CSEs.

Free-free emission is ruled out by McDonald et al. (2010) as an
important source of emission in their sample of giants in $\omega$ Cen. 
A MOLsphere could be due to many molecules but would most likely
manifest itself by the presence of water lines in the 6-8 $\mu$m region.  
This region is not covered by the LRS spectrum so it is impossible to
verify this idea directly.  McDonald et al. (2010) studied the effect
of using {\it Spitzer} IRS data and found that no reasonable combination
of column density and temperature could reproduce the flatness of their spectra.
For red supergiants, which are much more luminous that RGB stars, but
that have similar effective temperatures than the stars under study,
Verhoelst et al. (2006, 2009) found that a MOLsphere alone can not
explain the excess emission in the mid-IR and that a additional source
of opacity was needed.

\begin{figure*}[H]

\begin{minipage}{0.49\textwidth}
\resizebox{\hsize}{!}{\includegraphics[angle=-0]{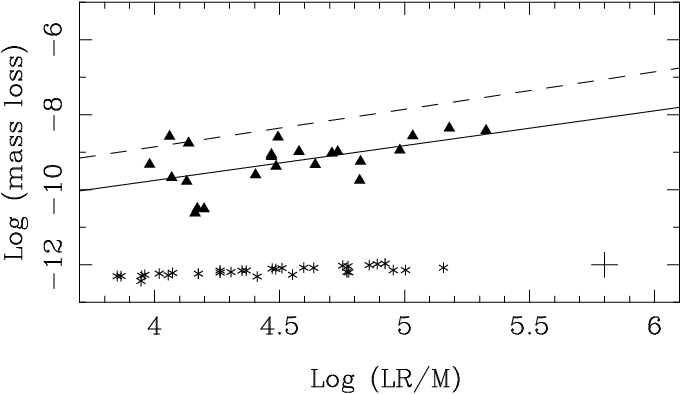}} 
\end{minipage}
\begin{minipage}{0.49\textwidth}
\resizebox{\hsize}{!}{\includegraphics[angle=-0]{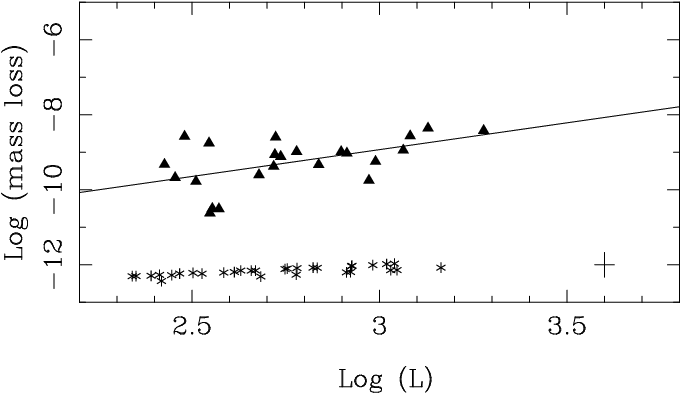}} 
\end{minipage}
 
\caption[]{ 
Mass-loss rate plotted against $LR/M$ (for a mass of 1 \msol) and $L$.
Stars for which no mass loss could be detected (an optical depth of 10$^{-5}$) are plotted as crosses.
The solid lines indicate least squares fits to the data (see Table~\ref{Tab-fits}), while the dashed line represents Reimers law with $\eta = 0.35$.
The cross in the lower right corner indicates a typical error bar. 
}
\label{fig-ML} 
 
\end{figure*}

\section{Discussion}

\subsection{Mass loss}

Reimers law represents the Reimers (1975) mass-loss rate formula for red giants given by: 
\begin{displaymath}
\dot{M} = \eta \cdot  4 \cdot 10^{-13} \; (\frac{L \cdot R}{M})^{\gamma} \,\,\, (M_{\odot}\,{\rm yr}^{-1}),
\end{displaymath}
 (with $\gamma=1$ and $\eta=1$, and $L$, $R$ and $M$ in solar units) 
which, interestingly, Kudritzki \& Reimers (1978) updated to $\dot{M} = (5.5 \pm 1) \cdot 10^{-13} \; \frac{L \cdot R}{M}$ (\msolyr), 
i.e. $\eta \approx 1.4$,  by considering the mass loss in $\alpha$ Her, $\alpha$ Sco, and $\delta^2$ Lyr.

The left-hand panel of Fig.~\ref{fig-ML} shows the results of the current work, {\it assuming} a mass of 1.1 \msol\ 
for all stars. The PARAM tool\footnote{http://stev.oapd.inaf.it/cgi-bin/param} (da Silva et al. 2006) was used to find 
that this is the typical mass of a star in the sample.
An unweighted least squares fit gives $\log \dot{M}$ (\msolyr) $= (0.9 \pm 0.3) \log (L \; R/M) + (-13.4 \pm 1.3)$.
The slope is consistent with unity, but the coefficient ($3.7 \cdot 10^{-14}$) is a factor of ten lower than in Reimers law.
When $\gamma$ is fixed to unity, the coefficient becomes $1.8 \cdot 10^{-14}$ ($\eta \approx 0.04$) with an error of a factor of 3.4.
The right-hand panel shows a fit of the mass-loss rate versus luminosity
$\log \dot{M} $ (\msolyr) $ = (1.4 \pm 0.4) \log L + (-13.2 \pm 1.2)$.
Similar plots were made with the sample divided into K- and M-giants, according to Hipparcos variability type and effective temperature, 
but no clear trends were found.
Fits against radius and effective temperature have also been made, and the results are compiled in Table~\ref{Tab-fits}.
The best-fit relation is obtained when the mass-loss rate is fitted against ($\log$) stellar radius.

Linear fits using two variables were also tested (Table~\ref{Tab-fits}), which resulted in lower $\chi^2$ but, according to 
the Bayesian information criterion (Schwarz 1978) where 
${\rm BIC} = \chi^2 + (p + 1)\; \ln (n)$ ($p$ is the number of free parameters, and $n$ the number of data points), 
this is not significant as the BICs are larger. 

These trends are in agreement with Catelan (2000), who in his
appendix also presents several simple fitting formula that fit the
data equally well.  In the case of the fit against radius (his Eq. A3),
he finds a slope of 3.2 (no error given), while we find a similar value of 2.6 $\pm$ 0.7. 
For a radius of 100 \rsol, Catelans' formula gives a mass-loss rate of
$3.0 \cdot 10^{-9}$, while we find $2.2 \cdot 10^{-9}$ \msolyr.

Among the 54 stars, 23 stars are found to have a significant infrared excess, 
which is interpreted as evidence for mass loss.  The most luminous star with 
$L= 1860\;\lsol$ is found to have mass loss, while none of the 5 stars with 
$L < 262\;\lsol$ show evidence for mass loss.
In the range 265 $< L < 1500 \;\lsol$, 22 stars out of 48 show mass loss,
which supports the notion of episodic mass loss proposed by Origlia et al. (2007).
They also find a shallower slope of $\gamma$ = 0.4, which the current data does not support. 
Catelan (2000) quotes $\gamma$ = 1.4.

The sample selection discussed in Sect.~2 involved imposing a lower limit of $(V-I)_0 > 1.5$,
which could lead to a bias favouring mass-losing stars.
To verify this, the $(V-I)$ colour predicted by the RT model of the mass-losing stars 
was compared to the colour of a model with the optical depth fixed to zero.  
Even for the star with the highest mass-loss rate, the model without mass loss is only 
bluer by 0.003 magnitudes.
This implies that the selection based on $(V-I)$ has no consequences for the 
statistics of the number and fraction of mass-losing stars in the sample.

\begin{table*}

\caption{Linear least squares fits to the mass-loss rates. 
First entries are fits to the current sample (see Fig.~\ref{fig-ML}).
Last three entries includes literature mass-loss rates from modelling chromospheric lines in GC RGB stars (see Fig.~\ref{fig-MLall}).
}

\centering
  \begin{tabular}{rrrrrrrrrr}
  \hline \hline

variable(s) &   zero point & slope  & slope &  $\chi^2$ & BIC  & \\ \hline

$\log R$                         & $-$13.76 $\pm$ 1.30 & 2.56 $\pm$ 0.74 & & 25.25 & 31.52 \\
$\log LR/M$                      & $-$13.43 $\pm$ 1.26 & 0.92 $\pm$ 0.27 & & 26.16 & 32.43 \\
$\log L$                         & $-$13.22 $\pm$ 1.23 & 1.42 $\pm$ 0.44 & & 26.88 & 33.16 \\
$\log T_{\rm eff}$                &  52.17 $\pm$ 20.3 & $-$17.2 $\pm$ 5.7 & & 28.10 & 34.73 \\

$\log R$ and $\log L$             & $-$13.90 $\pm$ 1.31 & 5.1 $\pm$ 3.4 & $-$1.5 $\pm$ 2.1 & 24.65 & 34.09 \\
$\log LR/M$ and $\log T_{\rm eff}$ & 19.3 $\pm$ 27.0 & 0.65 $\pm$ 0.35 & $-$8.8 $\pm$ 7.3 & 24.69 & 34.09 \\

 & \\
$\log L$                         & $-$11.87 $\pm$ 0.78 & 0.99 $\pm$ 0.27 & & 50.34 & 58.08 & unweighted OLS \\ 
$\log LR/M$                      & $-$11.83 $\pm$ 0.78 & 0.60 $\pm$ 0.17 & & 50.55 & 58.30 & unweighted OLS \\
$\log R$                         & $-$11.16 $\pm$ 0.87 & 1.22 $\pm$ 0.49 & & 57.67 & 65.21 & unweighted OLS \\
 & \\
$\log L$                         & $-$12.00 $\pm$ 0.94 & 1.04 $\pm$ 0.31 & &  & & BCES (OLS) \\ 

$\log LR/M$                      & $-$11.90 $\pm$ 0.92 & 0.62 $\pm$ 0.19 & &  & & BCES (OLS) \\

\hline

\hline
\end{tabular}

\label{Tab-fits}
\end{table*}

\begin{table*}

\renewcommand{\baselinestretch}{0.8}

\caption{GC data for mass-losing stars}
\footnotesize

\centering
  \begin{tabular}{rrrrrrrrrr}
  \hline \hline
 Cluster & Identifier & mass-loss rate & [Fe/H] & $T_{\rm eff}$ & $\log L$ & Reference \\ 
         &            &  (\msolyr)              &         &  (K)         & (\lsol) &   \\ \hline

M13 &   L72 & 2.8 \, 10$^{-09}$ & $-$1.54 & 4180. & 3.096 & Meszaros et al. (2009) \\
M13 &   L96 & 4.8 \, 10$^{-09}$ & $-$1.54 & 4190. & 3.010 & \\
M13 &  L592 & 2.6 \, 10$^{-09}$ & $-$1.54 & 4460. & 2.689 & \\
M13 &  L954 & 3.1 \, 10$^{-09}$ & $-$1.54 & 3940. & 3.329 & \\
M13 &  L973 & 1.6 \, 10$^{-09}$ & $-$1.54 & 3910. & 3.377 & \\
M15 &   K87 & 1.4 \, 10$^{-09}$ & $-$2.26 & 4610. & 2.708 & \\
M15 &  K341 & 2.2 \, 10$^{-09}$ & $-$2.26 & 4300. & 3.183 & \\
M15 &  K421 & 1.9 \, 10$^{-09}$ & $-$2.26 & 4330. & 3.207 & \\
M15 &  K479 & 2.3 \, 10$^{-09}$ & $-$2.26 & 4270. & 3.244 & \\
M15 &  K757 & 1.8 \, 10$^{-09}$ & $-$2.26 & 4190. & 3.195 & \\
M15 &  K969 & 1.4 \, 10$^{-09}$ & $-$2.26 & 4590. & 2.851 & \\
M92 & VII18 & 2.0 \, 10$^{-09}$ & $-$2.28 & 4190. & 3.208 & \\
M92 &   X49 & 1.9 \, 10$^{-09}$ & $-$2.28 & 4280. & 3.184 & \\
M92 &  XII8 & 2.0 \, 10$^{-09}$ & $-$2.28 & 4430. & 2.896 & \\
M92 & XII34 & 1.2 \, 10$^{-09}$ & $-$2.28 & 4660. & 2.570 & \\
NGC 2808 & 37872 & 1.1 \, 10$^{-09}$ & $-$1.14 & 4015. & 3.028 & Mauas et al. (2006)  \\
NGC 2808 & 47606 & 1.1 \, 10$^{-10}$ & $-$1.14 & 3839. & 3.218 & \\
NGC 2808 & 48889 & 3.8 \, 10$^{-09}$ & $-$1.14 & 3943. & 3.188 & \\
NGC 2808 & 51454 & 0.7 \, 10$^{-09}$ & $-$1.14 & 3893. & 3.177 & \\
NGC 2808 & 51499 & 1.2 \, 10$^{-09}$ & $-$1.14 & 3960. & 3.142 & \\
$\omega$ Cen & ROA159 & 1.1 \, 10$^{-09}$ & $-$1.72 & 4200. & 2.952 & Vieytes et al. (2011) \\
$\omega$ Cen & ROA256 & 6.0 \, 10$^{-09}$ & $-$1.71 & 4300. & 2.788 & \\
$\omega$ Cen & ROA238 & 3.2 \, 10$^{-09}$ & $-$1.80 & 4200. & 2.692 & \\
$\omega$ Cen & ROA523 & 5.0 \, 10$^{-10}$ & $-$0.65 & 4200. & 2.544 & \\
\\
47 Tuc &    V1 & 2.1 \, 10$^{-06}$ & $-$0.7 & 3623. & 3.683 & McDonald et al. (2011)  \\
47 Tuc &    V8 & 1.5 \, 10$^{-06}$ & $-$0.7 & 3578. & 3.554 & \\
47 Tuc &    V2 & 1.2 \, 10$^{-06}$ & $-$0.7 & 3738. & 3.482 & \\
47 Tuc &    V3 & 9.4 \, 10$^{-07}$ & $-$0.7 & 3153. & 3.473 & \\
47 Tuc &    V4 & 1.2 \, 10$^{-06}$ & $-$0.7 & 3521. & 3.415 & \\
47 Tuc &   V26 & 5.9 \, 10$^{-07}$ & $-$0.7 & 3500. & 3.405 & \\
47 Tuc &  LW10 & 4.2 \, 10$^{-07}$ & $-$0.7 & 3543. & 3.366 & \\
47 Tuc &   V21 & 4.9 \, 10$^{-07}$ & $-$0.7 & 3575. & 3.362 & \\
47 Tuc &   LW9 & 6.0 \, 10$^{-07}$ & $-$0.7 & 3374. & 3.343 & \\
47 Tuc &   V27 & 3.3 \, 10$^{-07}$ & $-$0.7 & 3374. & 3.330 & \\
47 Tuc & L1424 & 4.4 \, 10$^{-07}$ & $-$0.7 & 3565. & 3.327 & \\
47 Tuc &   A19 & 3.6 \, 10$^{-07}$ & $-$0.7 & 3526. & 3.321 & \\
47 Tuc &   x03 & 5.1 \, 10$^{-07}$ & $-$0.7 & 3816. & 3.215 & \\
47 Tuc &   V18 & 5.3 \, 10$^{-07}$ & $-$0.7 & 3692. & 3.113 & \\
47 Tuc &   V13 & 4.1 \, 10$^{-07}$ & $-$0.7 & 3657. & 3.012 & \\
$\omega$ Cen & 52111 & 5.7 \, 10$^{-08}$ & $-$1.62 & 3975. & 2.939 & McDonald et al. (2009) \\
$\omega$ Cen & 25062 & 1.2 \, 10$^{-07}$ & $-$1.83 & 4150. & 3.193 & \\
$\omega$ Cen & 43351 & 9.1 \, 10$^{-08}$ & $-$1.62 & 3895. & 3.002 & \\
$\omega$ Cen & 36036 & 7.8 \, 10$^{-08}$ & -2.05 & 3944. & 3.123 & \\
$\omega$ Cen & 42205 & 7.7 \, 10$^{-08}$ & $-$1.62 & 4110. & 2.985 & \\
$\omega$ Cen & 26025 & 7.5 \, 10$^{-08}$ & $-$1.68 & 4088. & 3.223 & \\
$\omega$ Cen & 45232 & 2.8 \, 10$^{-07}$ & $-$1.62 & 4276. & 3.279 & \\
$\omega$ Cen & 49123 & 1.9 \, 10$^{-07}$ & $-$1.62 & 3895. & 3.152 & \\
$\omega$ Cen & 48060 & 1.9 \, 10$^{-07}$ & $-$1.97 & 4117. & 3.219 & \\
$\omega$ Cen & 56087 & 1.8 \, 10$^{-07}$ & $-$1.92 & 4209. & 3.242 & \\
$\omega$ Cen & 41455 & 1.5 \, 10$^{-07}$ & $-$1.29 & 3966. & 3.127 & \\
$\omega$ Cen & 32138 & 1.4 \, 10$^{-07}$ & $-$1.87 & 4124. & 3.178 & \\
$\omega$ Cen & 37110 & 1.2 \, 10$^{-07}$ & $-$1.62 & 3981. & 2.957 & \\
$\omega$ Cen & 47153 & 1.0 \, 10$^{-08}$ & $-$1.62 & 4070. & 3.045 & \\
$\omega$ Cen & 48150 & 9.2 \, 10$^{-08}$ & $-$1.62 & 3956. & 3.226 & \\
$\omega$ Cen & 42302 & 8.3 \, 10$^{-08}$ & $-$1.62 & 4245. & 3.110 & \\
$\omega$ Cen & 39165 & 7.8 \, 10$^{-08}$ & $-$1.62 & 4144. & 3.002 & \\
$\omega$ Cen & 39105 & 6.5 \, 10$^{-08}$ & $-$0.85 & 3833. & 3.136 & \\
$\omega$ Cen & 33062 & 2.0 \, 10$^{-06 }$& $-$1.08 & 3534. & 3.342 & McDonald et al. (2011) \\
$\omega$ Cen & 44262 & 2.0 \, 10$^{-06}$ & $-$0.8 & 3427. & 3.209 & \\
$\omega$ Cen & 44277 & 1.0 \, 10$^{-06}$ & $-$1.37 & 3921. & 3.177 & \\
$\omega$ Cen & 55114 & 4.0 \, 10$^{-07}$ & $-$1.45 & 3906. & 3.164 & \\
$\omega$ Cen & 35250 & 6.0 \, 10$^{-07}$ & $-$1.06 & 3513. & 3.120 & \\
$\omega$ Cen & 42044 & 4.0 \, 10$^{-07}$ & $-$1.37 & 3708. & 3.102 & \\
NGC 362 & s02 & 1.7 \, 10$^{-06}$ & $-$1.16 & 3907. & 3.262 & Boyer et al. (2009) \\
NGC 362 & s03 & 7.1 \, 10$^{-07}$ & $-$1.16 & 4339. & 3.339 & \\
NGC 362 & s04 & 6.1 \, 10$^{-07}$ & $-$1.16 & 3823. & 3.191 & \\
NGC 362 & s05 & 9.3 \, 10$^{-07}$ & $-$1.16 & 4058. & 3.358 & \\
NGC 362 & s06 & 2.0 \, 10$^{-06}$ & $-$1.16 & 3962. & 3.492 & \\
NGC 362 & s07 & 1.3 \, 10$^{-06}$ & $-$1.16 & 3343. & 3.147 & \\
NGC 362 & s09 & 7.4 \, 10$^{-07}$ & $-$1.16 & 4226. & 3.286 & \\
NGC 362 & s10 & 4.8 \, 10$^{-07}$ & $-$1.16 & 3975. & 3.134 & \\

\hline
\end{tabular}

\label{Tab-GC}
\end{table*}

\medskip

Mass-loss rate estimates below the tip of the RGB exist mostly for stars in globular
clusters (GCs), derived by both modelling the SEDs, as in the present paper, and
modelling the chromospheric line profiles.
Data were collected from the literature and are reproduced in Table~\ref{Tab-GC}.
The first 24 entries are based on the modelling of the chromospheric line profiles, 
followed by the results of modelling the SEDs. For $\omega$ Cen, 
the values from McDonald et al. (2011) of the mass loss are preferred over those of McDonald et al. (2009) 
as they used \it Spitzer \rm IRS spectra as additional constraints in the fitting.

All the SED modelling was performed in a similar way, also using DUSTY.  
However, in the McDonald et al. (2009, 2011, 2011) and Boyer et al. (2009) papers, 
DUSTY was run using a mode assuming radiatively-driven winds ("density type = 3").
The output velocity of DUSTY was then scaled as
$(L/10^4)^{(1/4)} (\Psi/0.005)^{(1/2)} (\rho/3)^{(-1/2)}$, 
and the mass-loss rate as
$(L/10^4)^{(3/4)} (\Psi/0.005)^{(-1/2)} (\rho/3)^{(1/2)}$.
A dust-to-gas ratio of $\Psi = 0.005 \cdot 10^{\rm [Fe/H]}$ was assumed in these models, 
which corresponds to very low outflow velocities: Boyer et al. (2009) mention
values between 0.5 and 1.3 \ks\ for NGC 362, and McDonald et al. (2011) 
list values between 0.5 and 3.0 \ks\ for their sample in $\omega$ Cen. 

\begin{figure*}[H]

\begin{minipage}{0.49\textwidth}
\resizebox{\hsize}{!}{\includegraphics[angle=-0]{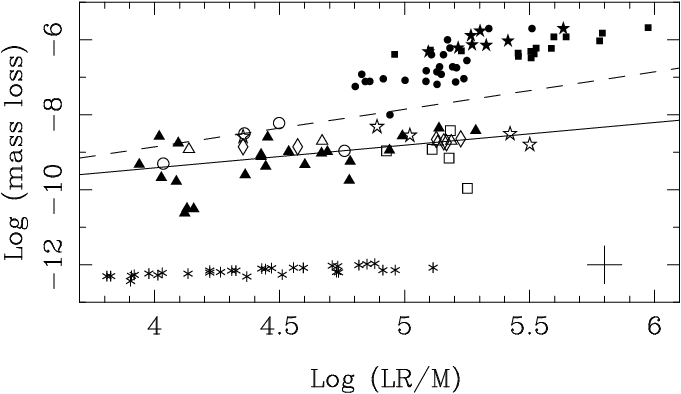}} 
\end{minipage}
\begin{minipage}{0.49\textwidth}
\resizebox{\hsize}{!}{\includegraphics[angle=-0]{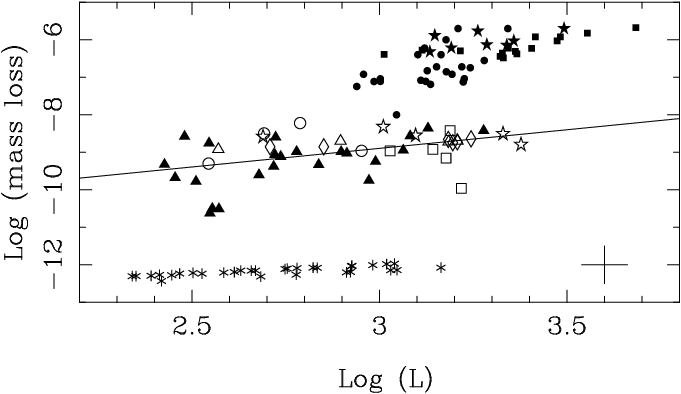}} 
\end{minipage}
 
\caption[]{ 
As Fig.~\ref{fig-ML}, but now for:
This work (filled triangles, and the stars without excess as crosses),
other data based on SED modelling:
47 Tuc (filled squares),
$\omega$ Cen (filled circles),
NGC 362 (filled stars),
and modelling of chromospheric lines:
M13 (open stars),
M15 (open diamonds),
M92 (open triangles), and
$\omega$ Cen (open circles).
The fit is to the present sample and the mass-loss rates from modelling the chromospheric lines, while the dashed line represents Reimers law with $\eta = 0.35$. 
The cross in the lower right corner indicates a typical error bar. 
} 
\label{fig-MLall} 
 
\end{figure*}

Figure~\ref{fig-MLall} is similar to Fig.~\ref{fig-ML} but with the literature data now added.
There is excellent agreement with the mass-loss rates based on the
modelling of the chromospheric activity, but an apparent discrepancy
with the data that are also based on modelling the SEDs. 
This discrepancy is discussed in more detail in Sect.~\ref{dustd}.

The results of unweighted least squares fits to the mass-loss rates derived in the present work and modelling of the chromospheric 
activity are reported in Table~\ref{Tab-fits} and shown in Figure~\ref{fig-MLall}. 
Adding the data from the chromospheric modelling leads to shallower, more accurately determined, slopes.

The use of unweighted least squares fits may be an oversimplification
as both the abscissa and ordinate have (different) error bars.  The error
in the mass-loss rate is difficult to quantify. Following the
discussion in Sect.~\ref{S-Results}, the internal fit error in the optical
depth (typically 5\%), the error in the optical depth due to
uncertainties in the stellar photosphere parameters (typically 50\%),
and the uncertainty in the expansion velocity and dust-to-gas ratio when
converting optical depth to mass-loss rates (assumed to be a factor of 2) are
added in quadrature. The last error term dominates.

The error in the mass-loss rate derived by modelling the chromospheric lines
is quoted to be a factor of two by Meszaros et al. (2009) and this is
taken to be the error for all mass-loss rates derived by modelling the
chromospheric lines. This means that the error bars along the ordinate are
quite similar for all objects used in the fitting.

For the stars studied in the present work, the error in luminosity was 
derived by adding in quadrature the internal fit error (Table~\ref{Tab-fit})
and the error in the parallax (Table~\ref{Tab-sample}).
The error in radius was calculated by taking the error in $L$ and a 70~K error in effective temperature.
The error in mass was assumed to be 0.1 \msol.
For the stars in the present sample, typical error bars in $\log L$ and $\log LR/M$ are 0.026, 
respectively 0.05 dex, and are shown in Fig.~\ref{fig-ML}.

For the stars in the GCs it turns out that the luminosities quoted in the literature 
have been determined from $V$ and $K$ magnitudes and bolometric corrections.
In addition there is the uncertainty in the distance modulus to the cluster. 
We assumed that an error bar of 0.15 in $M_{\rm bol}$ was representative.
The error in radius was calculated as before, while the error in mass was taken as 0.01\msol.
The typical error bars in $\log L$ and $\log LR/M$ are 0.06 and 0.07 dex, respectively.

The "bivariate correlated errors and intrinsic scatter" (BCES) method was used 
(Akritas \& Bershady 1996)\footnote{http://www.astro.wisc.edu/$^\sim$mab/archive/stats/stats.html}, 
assuming that the errors along the ordinate and abscissa are uncorrelated.
The results are very similar to those for the unweighted fitting, see Table~\ref{Tab-fits}, 
probably because the errors along the ordinate are similar 
for all data, and the errors along the abscissa are smaller than those in the ordinates.
In what follows, the results from the BCES method are used.

\subsection{Are the winds dust driven?}
\label{dustd}

Figure~\ref{fig-MLall} highlights an apparent discrepancy between the present work and 
the literature data also based on modelling the SEDs and that use the same numerical code.

The difference can be traced back to the way in which DUSTY is run, either with
"density type = 1" and assuming a $r^{-2}$ density law, the mass-loss rate derived from the optical depth, 
the assumed dust-to-gas ratio and expansion velocity (as in the present paper),
or, with "density type = 3" (which assumes the winds are driven by
radiation pressure on dust), and then taking the DUSTY output for the
mass-loss rate and expansion velocity, and applying scaling relations
(see above) that take into account the luminosity and dust-to-gas ratio.

As a test, DUSTY models with "density type = 3" were also run for the
present sample.  The (scaled) mass-loss rates were then compared to the
mass-loss rates from Table~\ref{Tab-fit} scaling those to the
predicted expansion velocities. Neglecting four outliers (with ratios above
100, which are among the five stars with the lowest mass-loss rates, 
hence optical depths that are probably particularly uncertain), 
the ratio of the  mass-loss rates are between 18 and 60, with a median of 38.

\begin{figure}[H]

\begin{minipage}{0.49\textwidth}
\resizebox{\hsize}{!}{\includegraphics[angle=-0]{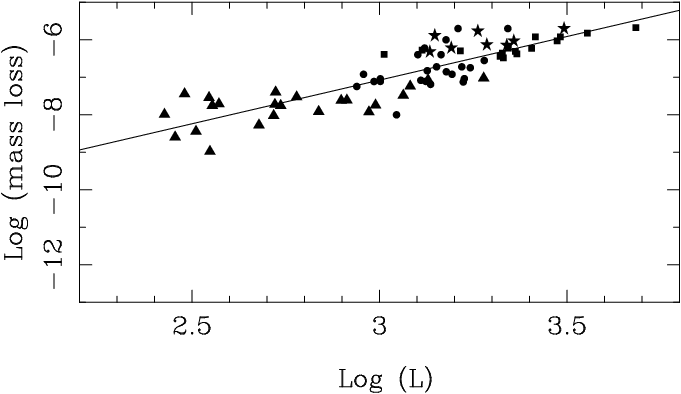}} 
\end{minipage}

\caption[]{ 
As Fig.~\ref{fig-ML}, with mass-loss rate plotted against luminosity, with 
mass-loss rates for the current sample calculated using DUSTY in
"density type = 3"
%
} 
\label{fig-ML-scaled} 
 
\end{figure}

Figure~\ref{fig-ML-scaled} shows the new results together with the
literature mass-loss rates from the SED modelling.  The results appear
now to be in much better agreement.  We note that this suggests that
there is a weak if any dependence of RGB mass loss on metallicity. 
The agreement between the mass-loss rates from the present modelling
and the chromospheric line emission of the GC stars indicates the same.

There is however a good reason to be cautious when using DUSTY in the dust-driven wind mode and 
that is that the ratio of radiative forces to gravitation may be less than unity in some cases.

Equation~15 in Elitzur \& Ivezi\'c (2001) derived this ratio to be $\Gamma =  45.8~Q_{\star} \sigma_{22} L_4 / M$, 
where $M$ is the stellar mass in solar units, $L_4$ the luminosity in units of 10$^4$ \lsol, 
$Q_{\star}$ is the Planck average at the stellar temperature of the efficiency coefficient for radiation pressure 
(Equation~4 in their paper), and $\sigma_{22}$ is the cross-section area per gas particle in units of $10^{-22}$ cm$^2$ 
(see Equation~5 in their paper).  The latter quantity can be written as (with Eq.~86 in their paper) 
$\sigma_{22} = 125.4 \, \Psi / (\rho \, a)$, with $\Psi$ the dust-to-gas ratio, 
$\rho$ the grain specific density in g cm$^{-3}$, and $a$ the grain size in micron.

In the optically thin case, $Q_{\star}$ can be taken as the sum of
absorption and scattering coefficients at the wavelength where the
stellar photosphere peaks.  For the typical effective temperatures
considered here, this is at about 1.6 micron, and $Q_{\star} \sim 2.2$
(for the iron dust properties discussed in Sect.~3).  With $\rho = 5.1$, 
$a = 0.15$ and $\Psi = 0.005$, $\sigma_{22}$ equals 0.82, and then, assuming 
$M = 1$, $\Gamma$ becomes 8.3 for $L_4 = 0.1$.  In this case, $\Gamma > 1$ 
and so the condition to drive the outflow by radiation pressure on dust is met. 
For the lowest luminosities in which we detect an excess, $L$ is $\sim$260 \lsol\ and so $\Gamma \sim 2$.

In the case of the GC stars from the literature, the majority of luminosities 
are above $L_4 = 1$ (see Fig.~\ref{fig-MLall}) but these authors assumed 
the dust-to-gas ratio to scale with metallicity, which results in very low values 
in the range 1/1200 to 1/5000 in the case of $\omega$ Cen (McDonald et al. 2011), and again, 
$\Gamma$ would be close to or even below unity.

The largest uncertainty in predicting $\Gamma$ may be in the calculation of the dust properties. 
As mentioned in the introduction, there are indications that metallic iron may be abundant and 
the fact that the infrared excess is featureless is one of them.

A value of $Q_{\star} = 2.2$ is much larger than that found for standard silicates or carbon dust 
(Table~1 in Elitzur \& Ivezi\'c 2001 lists 0.1 and 0.6, respectively, for 0.1 $\mu$m sized grains).
The adopted grain size is of importance. In the standard case, $Q/a = 14.6$~$\mu$m$^{-1}$, 
but this value decreases (and hence $\Gamma$ decreases) to 9.4 and 6.1 for $a = 0.1$ and 0.05~$\mu$m, respectively.
This effect can also lead to a larger $\Gamma$: $Q/a$ peaks at  21.6~$\mu$m$^{-1}$ for $a = 0.23~$$\mu$m .
The influence on grain size is largely due to the scattering contribution and this implies that even the assumption 
of isotropic scattering can play a role.
The assumption on the grain morphology is also important. If the DHS is used with a smaller maximum fraction 
of vacuum $Q/(a \rho)$ is also reduced: by a factor of 2 in the case of $f_{\rm max} = 0.4$ 
and a factor of 3.2 in the case of $f_{\rm max} = 0.0$ (i.e. compact spherical grains).

Elitzur \& Ivezi\'c (2001) derive in a similar way a more stringent
constraint on the minimal mass-loss rate (their Eq.~69), 
\mdot $\ga$ $3 \cdot 10^{-9} \; \frac{M}{Q_{\star} \; \sigma_{22}^2 L_4 T_{\rm k3}^{0.5}}$,
which corresponds to $2 \cdot 10^{-8}$ \msolyr\ for the above values for 
$M, Q_{\star}$, and $\sigma_{22}$, $L_4 = 0.1$, and $T_{\rm k3} = 1$, which is 
the kinetic temperature at the inner radius in units of 1000~K.
On the basis of this consideration. \it all \rm mass-loss rates in Table~\ref{Tab-fit} are below this critical value.

\subsection{Angular diameters}

Angular diameters are a prediction of the radiative transfer modelling
through the fitting of the luminosity and the effective temperature.
The values are listed in Table~\ref{Tab-fit}. 
The (limb darkened) angular diameters of some stars determined from interferometry
and the predicted values from the SED modelling are compared in Table~\ref{Tab-angd}.

\begin{table*}

\caption{Comparison of angular diameters.}

\centering
  \begin{tabular}{rrrrrrrrrr}
  \hline \hline
HIP    &   \multicolumn{4}{c}{$\theta$ and error  (mas)}            \\
       &   Bord\'e et al. (2002)       & Mozurkewich et al. (2003) & Richichi et al. (2009) & This paper (Table~\ref{Tab-fit}) \\ 
\hline

  4147  & 3.51 $\pm$ 0.037 &                   & 3.459 $\pm$ 0.006   &  3.85 $\pm$ 0.15 \\
 12107  & 3.11 $\pm$ 0.032 &                   &                     &  3.47 $\pm$ 0.15 \\
 32173  & 2.73 $\pm$ 0.029 &                   &                     &  3.03 $\pm$ 0.13 \\
 37300  & 3.28 $\pm$ 0.038 &                   &                     &  3.43 $\pm$ 0.13 \\
 41822  & 2.88 $\pm$ 0.031 &                   &                     &  2.58 $\pm$ 0.11 \\
 44857  & 2.67 $\pm$ 0.035 &                   &                     &  3.17 $\pm$ 0.13 \\
 46750  & 4.12 $\pm$ 0.046 &                   &                     &  4.59 $\pm$ 0.18 \\
 49005  & 2.18 $\pm$ 0.023 &                   &                     &  2.15 $\pm$ 0.08 \\
 53907  & 3.87 $\pm$ 0.041 &                   &                     &  3.95 $\pm$ 0.16 \\
 55687  & 3.57 $\pm$ 0.038 &                   &                     &  3.31 $\pm$ 0.12 \\
 56127  & 3.03 $\pm$ 0.034 &                   &                     &  3.41 $\pm$ 0.16 \\
 56211  &                  & 6.430 $\pm$ 0.069 &                     &  6.85 $\pm$ 0.27 \\
 60122  & 3.34 $\pm$ 0.038 &                   &                     &  3.60 $\pm$ 0.14 \\
 64607  & 2.29 $\pm$ 0.026 &                   &                     &  2.26 $\pm$ 0.11 \\
 67665  & 5.35 $\pm$ 0.059 &                   &                     &  5.64 $\pm$ 0.24 \\
 77661  & 3.40 $\pm$ 0.041 &                   &                     &  3.35 $\pm$ 0.12 \\
 82073  & 2.84 $\pm$ 0.029 &                   &                     &  2.80 $\pm$ 0.11 \\
 84835  & 2.60 $\pm$ 0.027 &                   &                     &  2.67 $\pm$ 0.11 \\
 87833  &                  & 9.860 $\pm$ 0.128 &                     & 10.55 $\pm$ 0.38 \\
 88122  & 2.42 $\pm$ 0.027 &                   &                     &  2.07 $\pm$ 0.08 \\
 106140 &                  & 4.521 $\pm$ 0.047 &                     &  4.89 $\pm$ 0.19 \\
 112716 & 5.12 $\pm$ 0.053 &                   & 5.008 $\pm$ 0.136   &  4.98 $\pm$ 0.19 \\

\hline
\end{tabular}

\label{Tab-angd}
\end{table*}

The largest overlap is that with Bord\'e et al. (2002), which is based on the results by Cohen et al. (1999). 
The overall agreement is good, leading to
$(\theta_{\rm Borde\; et\; al} - \theta_{\rm present\; work})/(\sqrt{\sigma_{\rm Borde\; et\; al}^2 + \sigma_{\rm present\; work}^2})$ 
of $-1.0 \pm 1.7$.
There are 2 stars where the difference is more than 3$\sigma$, HIP 44857 and HIP 88122. 
In the latter case, this is due to the different adopted effective temperatures,
4000~K in the present paper, and 3690~K in Bord\'e et al., which is
the largest difference in temperature among the 19 stars in common.

The overlap with Mozurkewich et al. (2003) occurs for three sources, for which the agreement is satisfactory. 
These authors obtained their data mostly at 550 and 800 nm and 
therefore the limb-darkening correction is relatively large.
The main difference is found for HIP 87833, for which Mozurkewich et al. quote 
angular diameters based on the infrared flux method of 10.450 (Bell \& Gustafsson 1989) 
and 10.244 mas (Blackwell et al. 1990), which are in excellent agreement with the present work.
Finally, two stars overlap with Richichi et al. (2009), and the agreement is satisfactory.

\section{Summary and conclusions}

We have selected a sample of 54 nearby RGB stars, and constructed and modelled their SEDs with a dust radiative transfer code.
In about half of the stars, the SEDs were statistically better fitted when we included a (featureless) dust component.
The lowest luminosity is which a significant excess was found is 267 \lsol\ (HIP 64607), which is 
lower than for any other star (as far as I am aware).

The results are compared with mass-loss rates estimated for RGB stars in GCs, 
based on modelling chromospheric line emission and the dust excess.
There is good agreement with the former sample, and fits of the mass-loss rate against luminosity and 
$(L\;R/M)$ (Reimers law) are presented by combining the present work with the mass-loss rates from the chromospheric modelling.
The derived slope is shallower then in Reimers law and with a larger constant.
The comparison with the literature values based on modelling the dust excess has led to an interesting observation.
This is that there is a significant difference among (and dependence on luminosity for) the mass-loss rates derived 
when running the DUSTY code in 
"density type = 1" and assuming a $r^{-2}$ density law, and then 
deriving the mass-loss rate from the optical depth,  an assumed 
dust-to-gas ratio, and expansion velocity (as in the present paper),
or, with "density type = 3" (that assumes the winds are driven by
radiation pressure on dust), and then taking the DUSTY output for the
mass-loss rate and expansion velocity, and applying scaling relations 
that take into account the luminosity and dust-to-gas ratio (as done in the literature).
The origin of the difference is unclear, but it turns out that the
ratio of radiative forces to gravitation ($\Gamma)$ could be smaller than unity under
certain conditions for the RGB stars we consider, for low
luminosity and/or low dust-to-gas ratios (as assumed in the GC stars). 
In addition, the details of the dust that forms (iron dust has
been assumed in the nearby RGB stars, as it is found to be the dominant
dust species in the GC sample), and the typical grain size and morphology could 
play a crucial role in whether $\Gamma$ is larger than unity in any given star.
This might explain why only 22 out of 48 RGB stars with 265 $< L < 1500 \;\lsol$ have an infrared excess.
The condition $\Gamma >1$ may only be fulfilled for 50\% of the time
(or, instantaneously, in 50\% of the stars) in this luminosity range when certain conditions are met.
What these conditions or triggers are remains unclear: they could be related to either pulsation or convection 
or be more indirect as for binarity or magnetic fields.
Determining the outflow velocity of the wind for the stars that show an excess would be helpful because this would 
not only remove the assumption of a constant velocity of 10 \ks\ for all objects, 
but also test the predictions of the dust-driven wind theory.

To investigate the implications of the mass-loss rate formula derived
here, we compared the predicted mass loss on the RGB to the recent determination 
from asteroseismology for the cluster NGC 6791 (Miglio et al. 2012), who derive 
a mass loss of $\Delta M$ = 0.09 $\pm$ 0.03 (random) $\pm$ 0.04 (systematic) \msol.
Evolutionary tracks were used from the same dataset (Bertelli et al. 2008\footnote{http://stev.oapd.inaf.it/YZVAR/}) 
and with the same composition (Z=0.04, Y=0.33) as used by Miglio et al.
For a star with initial mass 1.2 \msol\ (Miglio et al. determine the mass on the RGB to be 1.23 \msol), 
the mass lost on the RGB was calculated when the luminosity was above 250 \lsol\ (the effect of
including the entire RGB is negligible) for different mass-loss recipes (see Table~\ref{Tab-ML}).
The first entry is for Reimers law with a scaling of $\eta = 0.35$, which gives a total mass lost on the RGB 
close to the observed value.
The following rows provide the predictions from the best-fit relations derived in
this paper with $L$ and $LR/M$ as variables and varying the slope and zero point by their respective 1$\sigma$ error bars. 
Both type of relations can equally well result in the observed mass lost, 
for a slope and/or zero point slightly larger than the best-fit value.
As for the scaling of the Reimers law, the following relation would fit the observed mass loss of 0.09 $\pm 0.05$ \msol\ equally well:
$\dot{M} = \eta_1 \cdot  1.25 \cdot 10^{-12} \; (\frac{L \cdot R}{M})^{0.6}$ with $\eta_1 = 7.5 \pm 4$,
and
$\dot{M} = \eta_2 \cdot  1.00 \cdot 10^{-12} \; (L)^{1.0}$  with $\eta_2 = 9 \pm 5$.
That the scaling factors are larger than unity would suggest, in the framework of the dust modelling, 
that the expansion velocities and/or dust-to-gas ratios (or even the dust opacities) different from those assumed. 

The table also lists the predicted mass-loss rate at a luminosity of
1000 \lsol, and for the models that predict the observed total mass loss, 
this mass-loss rate is about  8 $\cdot 10^{-9}$ \msolyr.
Comparing this to Figures~\ref{fig-ML} and \ref{fig-ML-scaled} gives
independent evidence that the mass-loss rates in Table~\ref{Tab-GC} based on DUSTY using
"density type = 3" appear to be too large by an order of magnitude.

\begin{table*}

\caption{Mass lost on the RGB for a 1.2 \msol\ star
}

\centering
  \begin{tabular}{rrrrrrrrrr}
  \hline \hline

 relationship & $\Delta M$ & $\dot{M}$ (at L=1000\lsol) & remark  \\ 
              &  (\msol)   & 10$^{-9}$ \msolyr          & \\
\hline

$1.0 \cdot \log (LR/M)$ $-$13.097 & 0.090 &  6.9 & Reimers law $\eta$ = 0.20 \\

$0.6 \cdot \log (LR/M)$ $-$11.9   & 0.012 &  1.1  \\
$0.8 \cdot \log (LR/M)$ $-$11.9   & 0.127 & 11.1 \\
$0.4 \cdot \log (LR/M)$ $-$11.9   & 0.001 &  0.11 \\
$0.6 \cdot \log (LR/M)$ $-$11.0   & 0.096 &  9.1 \\
$0.6 \cdot \log (LR/M)$ $-$12.8   & 0.002 &  0.14 \\

$1.0 \cdot \log (L)$ $-$12.0   & 0.010 &  1.0 \\
$1.3 \cdot \log (L)$ $-$12.0   & 0.087 &  8.0 \\
$0.7 \cdot \log (L)$ $-$12.0   & 0.001 &  0.13 \\
$1.0 \cdot \log (L)$ $-$11.1   & 0.083 &  8.0 \\
$1.0 \cdot \log (L)$ $-$12.9   & 0.001 &  0.12 \\

\hline
\end{tabular}

\label{Tab-ML}
\end{table*}

\acknowledgements{  
MG would like to thank Dr. Iain McDonald for the very fruitful discussion regarding the input to DUSTY 
and commenting on a draft version of the paper, and the referee for suggesting additional checks, and pointing out the PARAM website.
This research has made use of the SIMBAD database, operated at CDS, Strasbourg, France. 
}

{}

\begin{appendix}
\section{Fits to the SEDs}

All fits are shown here.

\

\newpage

\

\begin{figure*}[H]

\begin{minipage}{0.42\textwidth}
\resizebox{\hsize}{!}{\includegraphics[angle=-0]{hip4147_sed.ps}} 
\end{minipage}
\begin{minipage}{0.42\textwidth}
\resizebox{\hsize}{!}{\includegraphics[angle=-0]{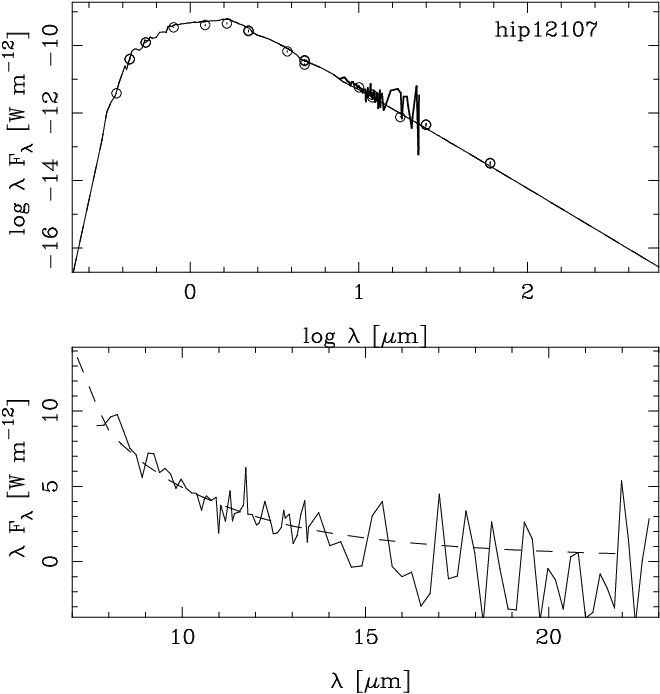}} 
\end{minipage}
 
\begin{minipage}{0.42\textwidth}
\resizebox{\hsize}{!}{\includegraphics[angle=-0]{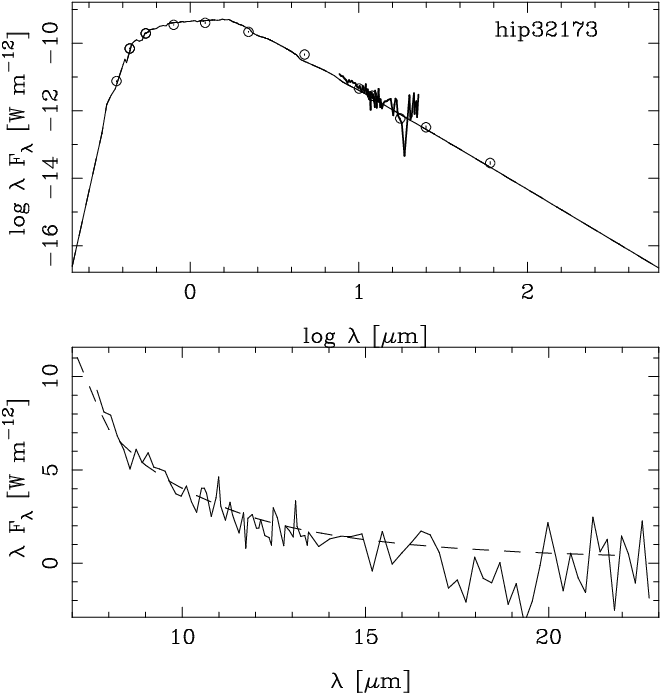}} 
\end{minipage}
\begin{minipage}{0.42\textwidth}
\resizebox{\hsize}{!}{\includegraphics[angle=-0]{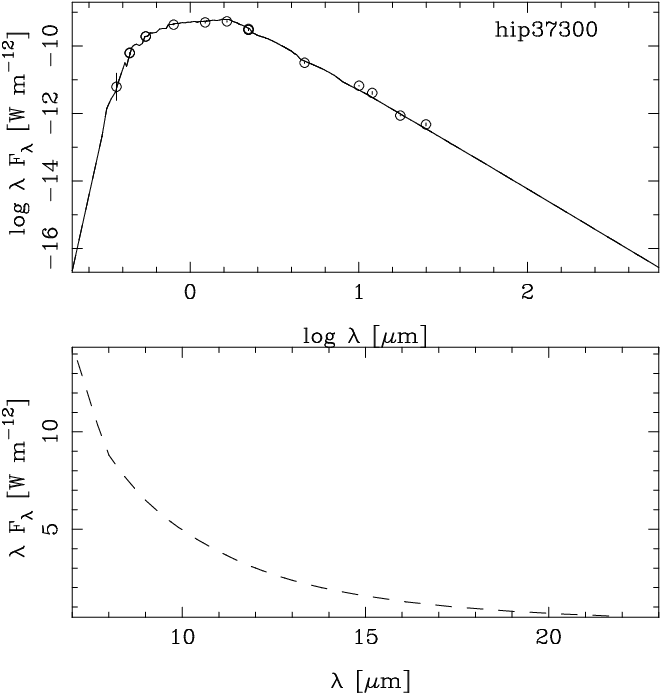}} 
\end{minipage}
 
\begin{minipage}{0.42\textwidth}
\resizebox{\hsize}{!}{\includegraphics[angle=-0]{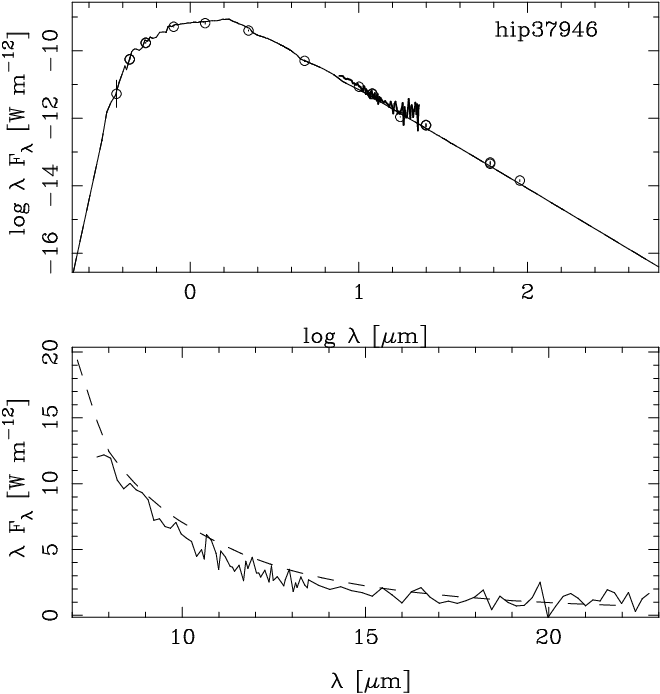}} 
\end{minipage}
\begin{minipage}{0.42\textwidth}
\resizebox{\hsize}{!}{\includegraphics[angle=-0]{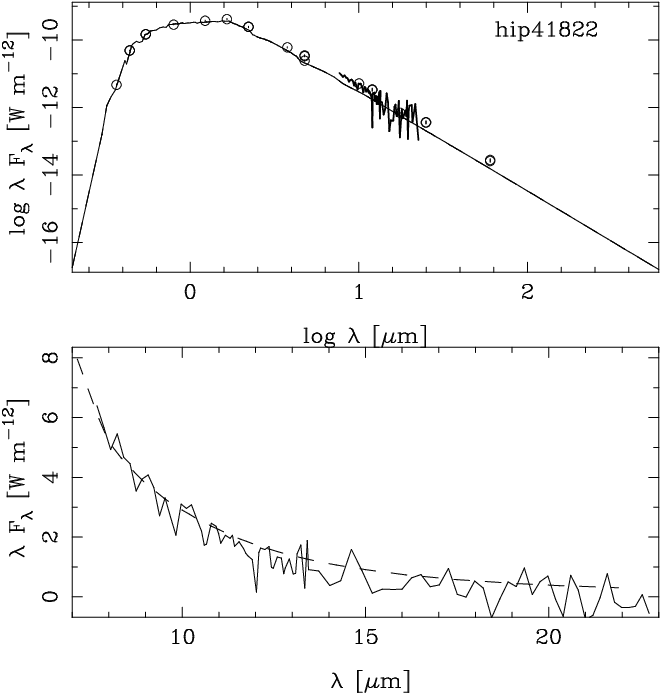}} 
\end{minipage}
 
\caption[]{ 
Fits to the SED (top panel) and IRAS LRS spectra (lower panel).
In the top panel, the solid line indicates the best fit, the dashed line the model without mass loss 
(in many cases the two models overlap and are indistinguishable).
The observed photometry is inidicated by the circles, and error bars are also plotted, but typically are much smaller than the symbol size.
In the lower panel, the best-fit model is indicated by the dashed line, and the LRS spectrum by the solid line. 
Sometimes no LRS spectrum was available.
} 
\label{Fig-SED-App} 

\end{figure*}

\begin{figure*}[H]

\begin{minipage}{0.42\textwidth}
\resizebox{\hsize}{!}{\includegraphics[angle=-0]{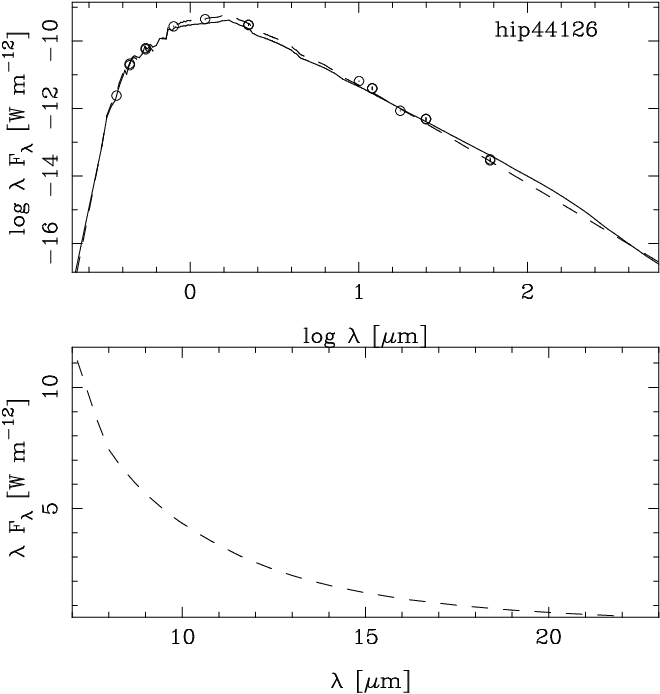}} 
\end{minipage}
\begin{minipage}{0.42\textwidth}
\resizebox{\hsize}{!}{\includegraphics[angle=-0]{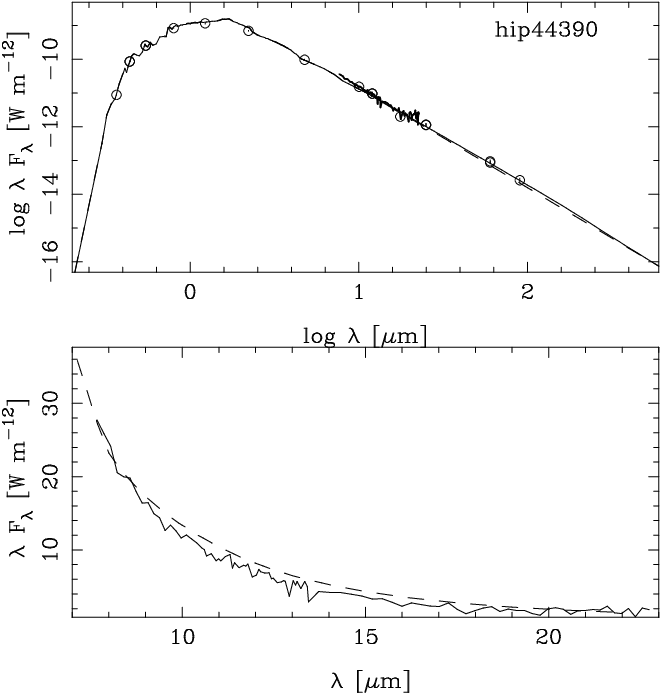}} 
\end{minipage}
 
\begin{minipage}{0.42\textwidth}
\resizebox{\hsize}{!}{\includegraphics[angle=-0]{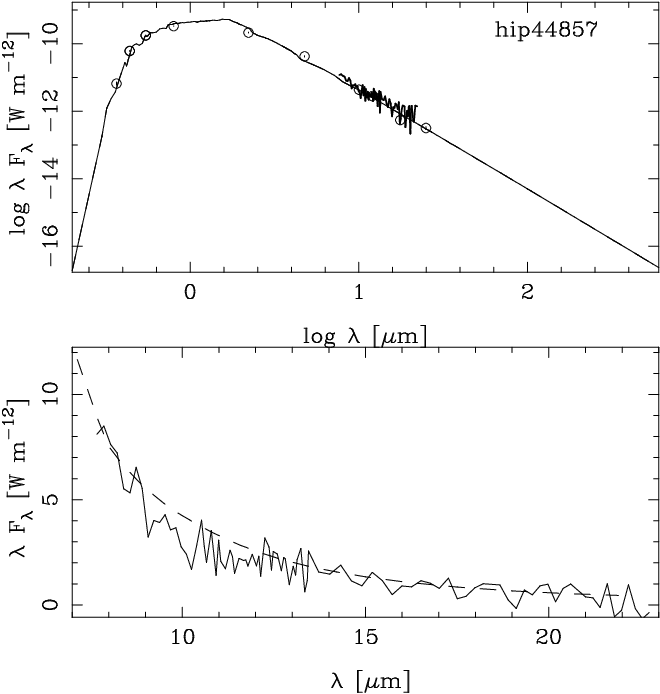}} 
\end{minipage}
\begin{minipage}{0.42\textwidth}
\resizebox{\hsize}{!}{\includegraphics[angle=-0]{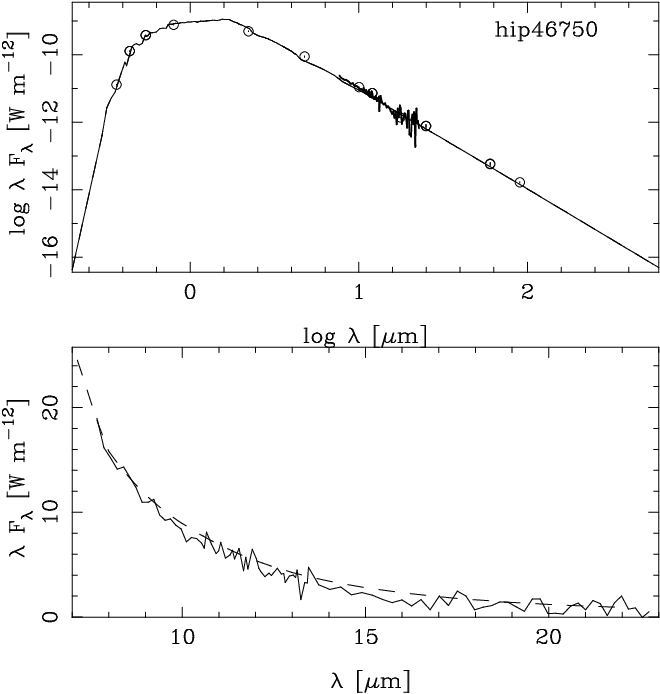}} 
\end{minipage}
 
\begin{minipage}{0.42\textwidth}
\resizebox{\hsize}{!}{\includegraphics[angle=-0]{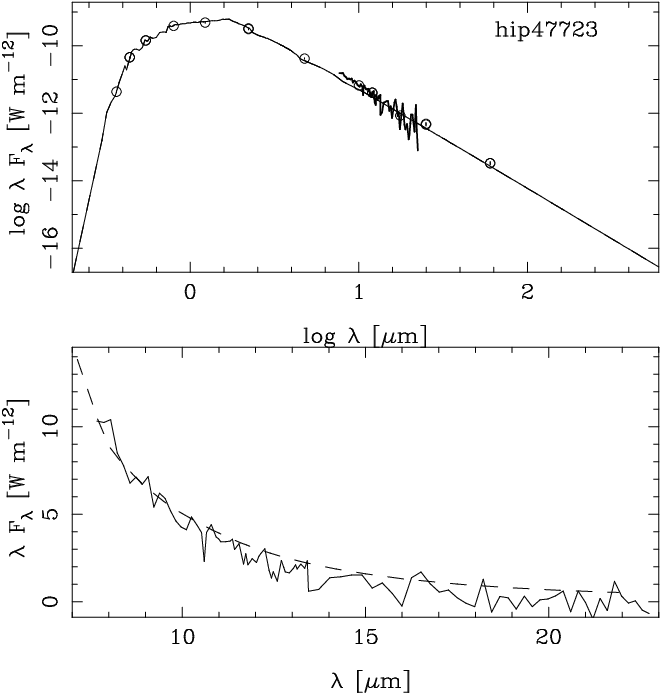}} 
\end{minipage}
\begin{minipage}{0.42\textwidth}
\resizebox{\hsize}{!}{\includegraphics[angle=-0]{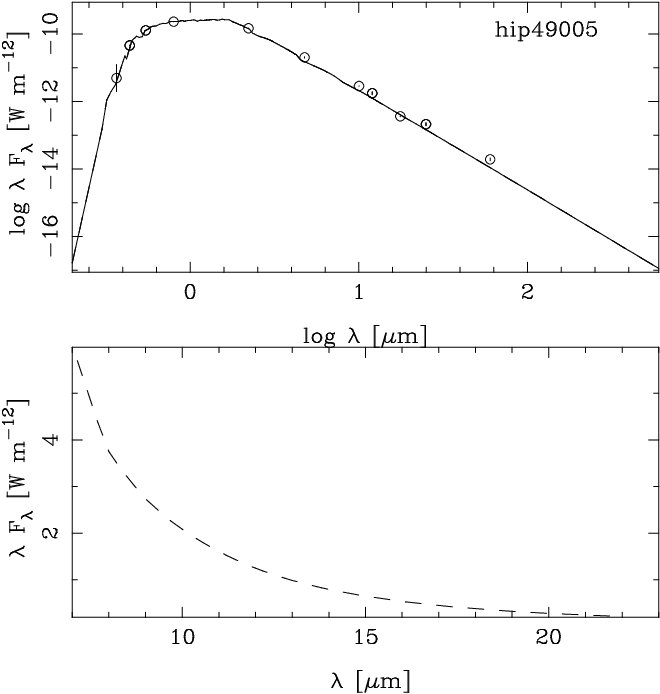}} 
\end{minipage}
 
\caption{Continued. } 

\end{figure*}

\begin{figure*}[H]

\begin{minipage}{0.42\textwidth}
\resizebox{\hsize}{!}{\includegraphics[angle=-0]{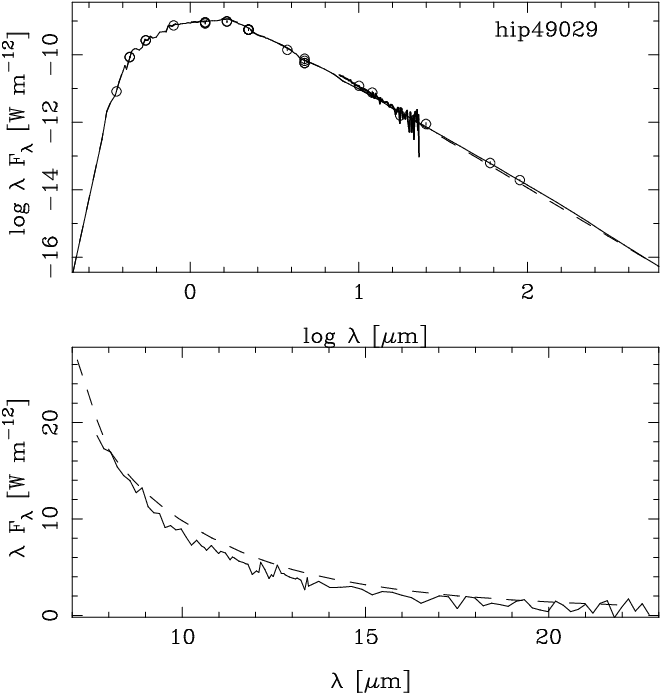}} 
\end{minipage}
\begin{minipage}{0.42\textwidth}
\resizebox{\hsize}{!}{\includegraphics[angle=-0]{hip52366_sed.ps}} 
\end{minipage}
 
\begin{minipage}{0.42\textwidth}
\resizebox{\hsize}{!}{\includegraphics[angle=-0]{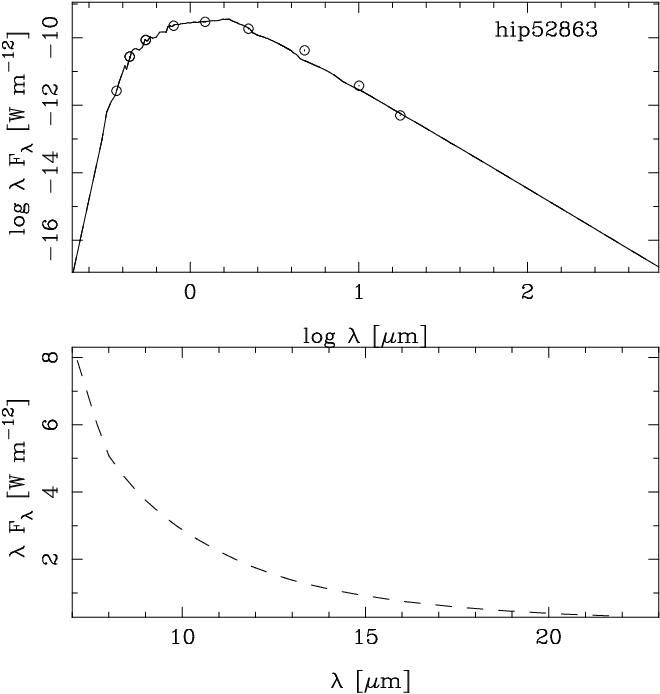}} 
\end{minipage}
\begin{minipage}{0.42\textwidth}
\resizebox{\hsize}{!}{\includegraphics[angle=-0]{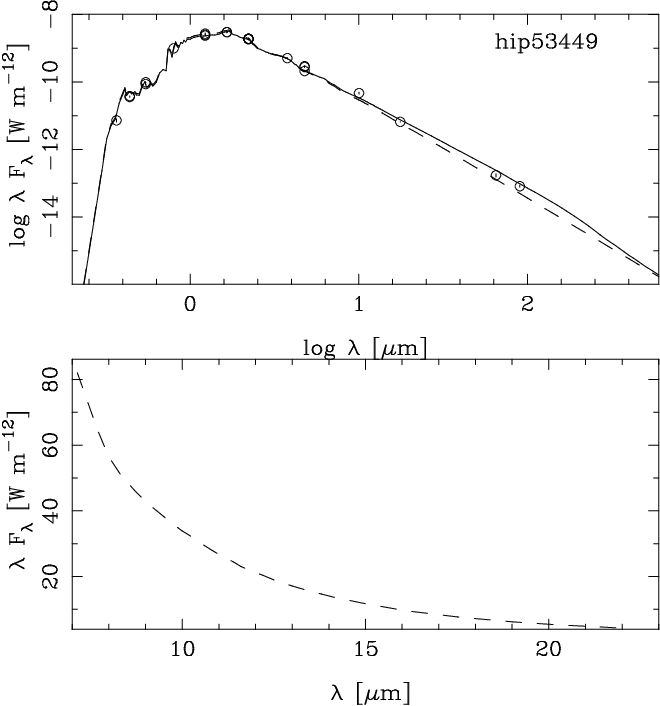}} 
\end{minipage}
 
\begin{minipage}{0.42\textwidth}
\resizebox{\hsize}{!}{\includegraphics[angle=-0]{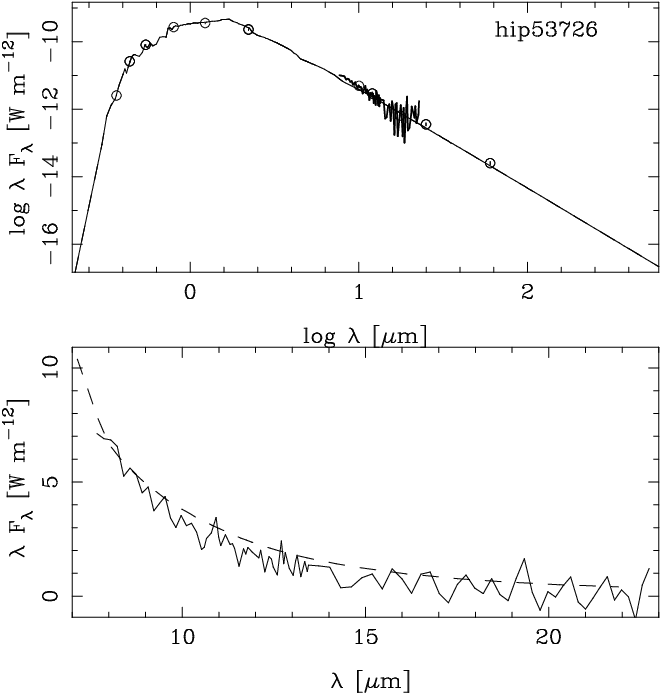}} 
\end{minipage}
\begin{minipage}{0.42\textwidth}
\resizebox{\hsize}{!}{\includegraphics[angle=-0]{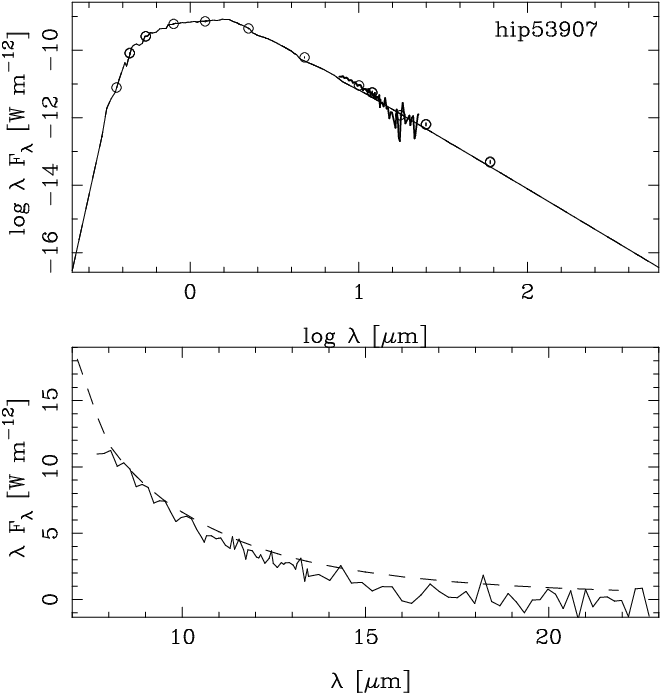}} 
\end{minipage}
 
\caption{Continued. } 

\end{figure*}

\begin{figure*}[H]

\begin{minipage}{0.42\textwidth}
\resizebox{\hsize}{!}{\includegraphics[angle=-0]{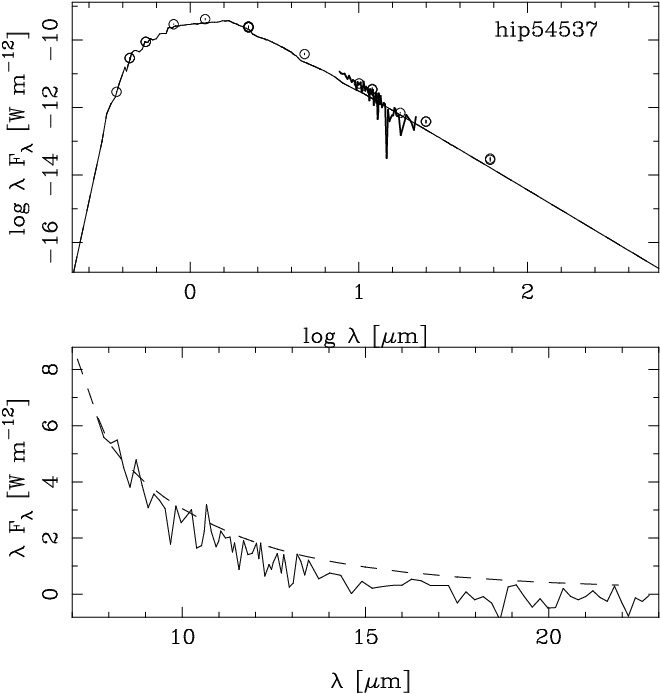}} 
\end{minipage}
\begin{minipage}{0.42\textwidth}
\resizebox{\hsize}{!}{\includegraphics[angle=-0]{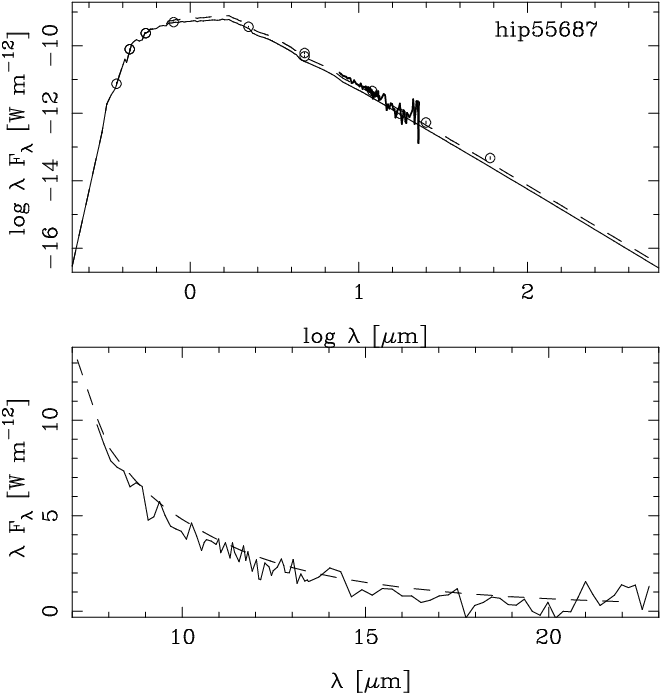}} 
\end{minipage}
 
\begin{minipage}{0.42\textwidth}
\resizebox{\hsize}{!}{\includegraphics[angle=-0]{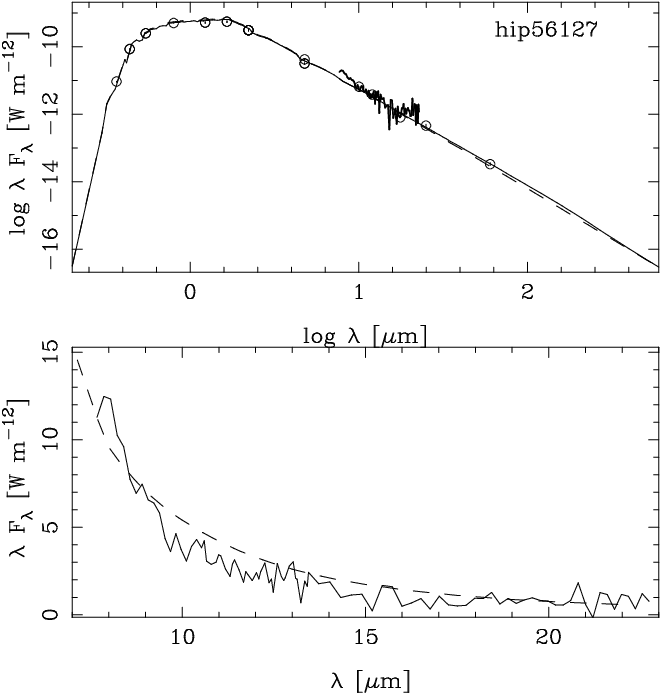}} 
\end{minipage}
\begin{minipage}{0.42\textwidth}
\resizebox{\hsize}{!}{\includegraphics[angle=-0]{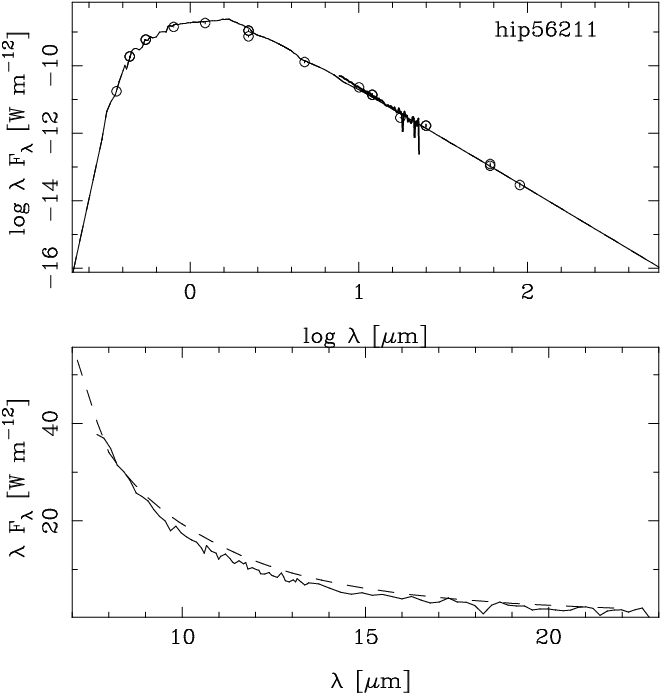}} 
\end{minipage}
 
\begin{minipage}{0.42\textwidth}
\resizebox{\hsize}{!}{\includegraphics[angle=-0]{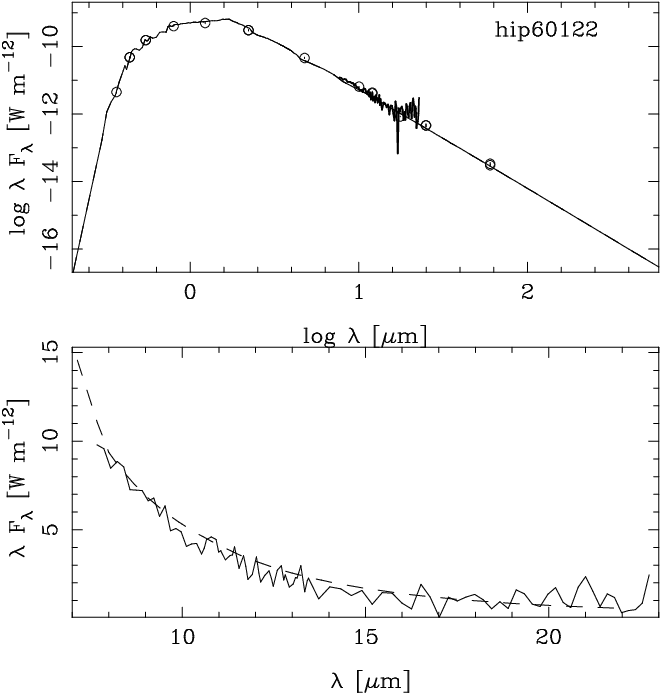}} 
\end{minipage}
\begin{minipage}{0.42\textwidth}
\resizebox{\hsize}{!}{\includegraphics[angle=-0]{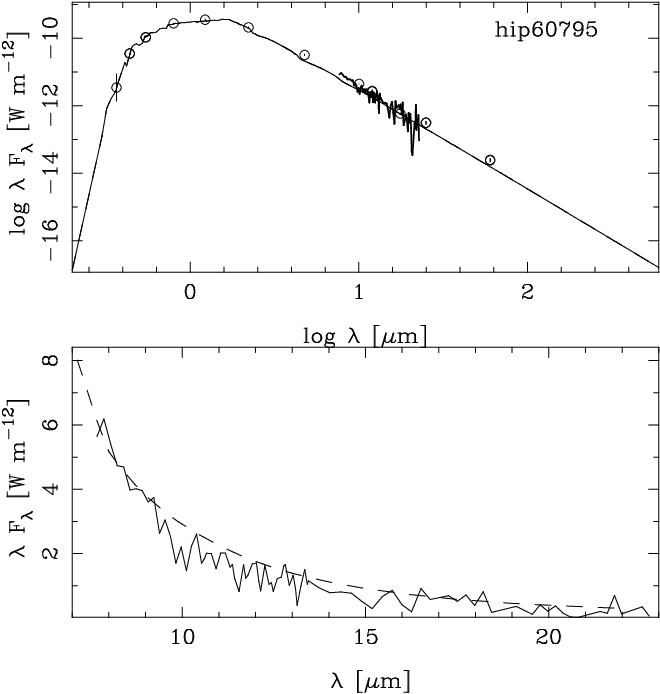}} 
\end{minipage}
 
\caption{Continued. } 

\end{figure*}

\begin{figure*}[H]

\begin{minipage}{0.42\textwidth}
\resizebox{\hsize}{!}{\includegraphics[angle=-0]{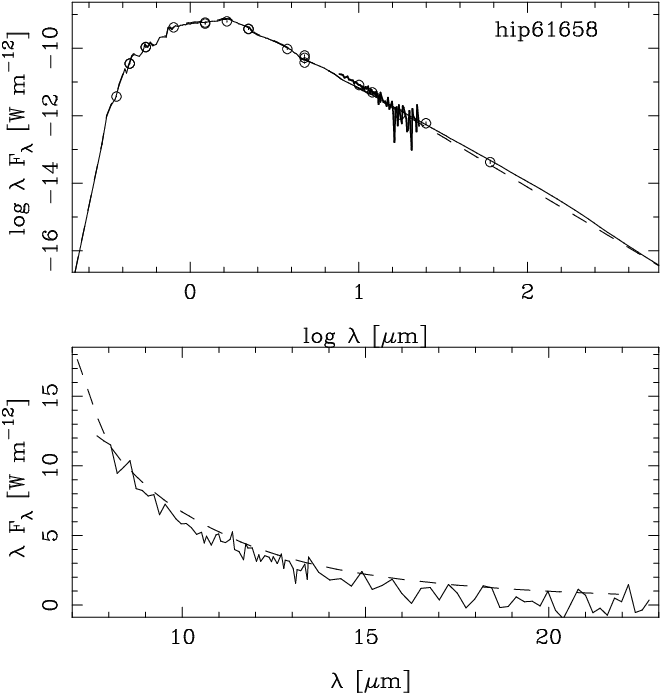}} 
\end{minipage}
\begin{minipage}{0.42\textwidth}
\resizebox{\hsize}{!}{\includegraphics[angle=-0]{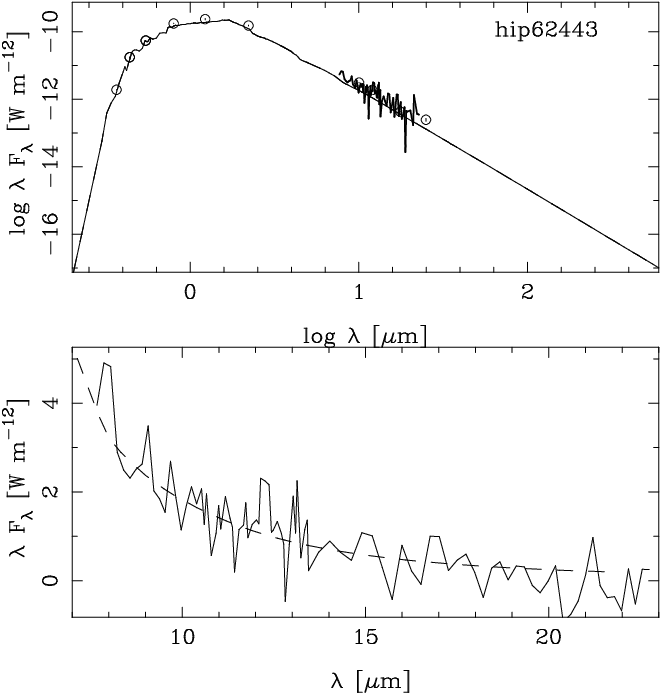}} 
\end{minipage}
 
\begin{minipage}{0.42\textwidth}
\resizebox{\hsize}{!}{\includegraphics[angle=-0]{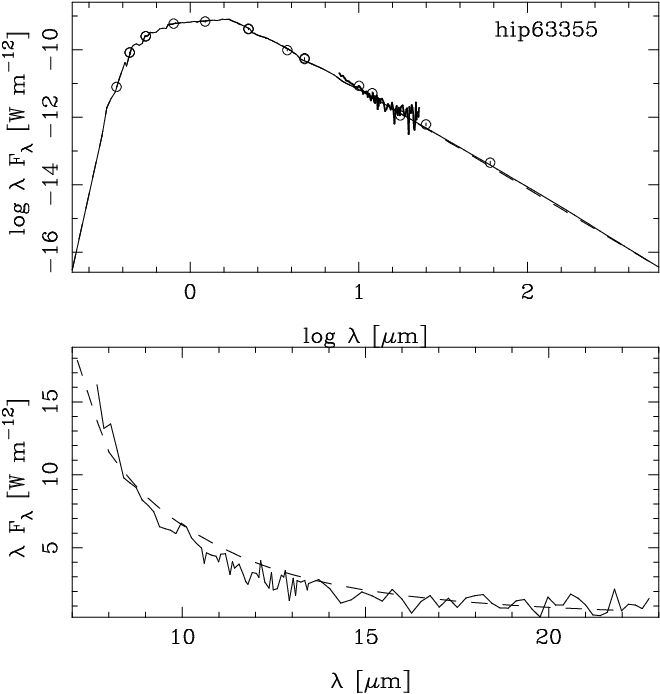}} 
\end{minipage}
\begin{minipage}{0.42\textwidth}
\resizebox{\hsize}{!}{\includegraphics[angle=-0]{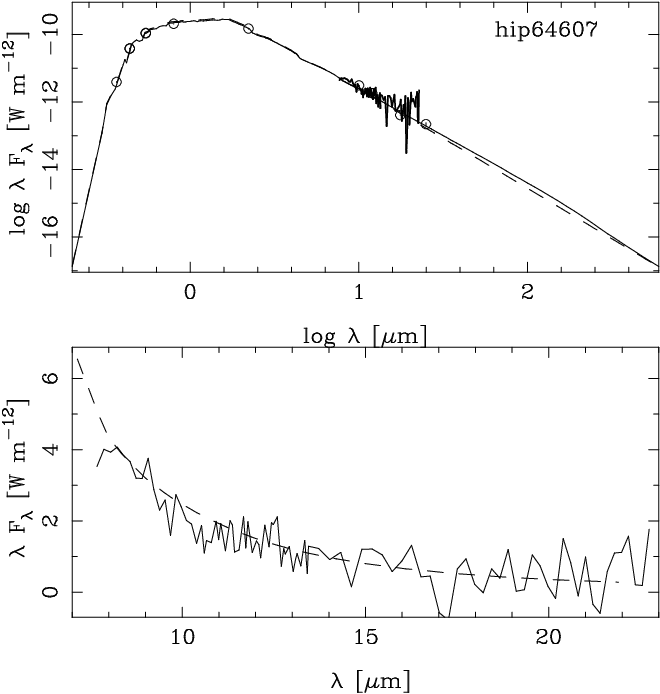}} 
\end{minipage}
 
\begin{minipage}{0.42\textwidth}
\resizebox{\hsize}{!}{\includegraphics[angle=-0]{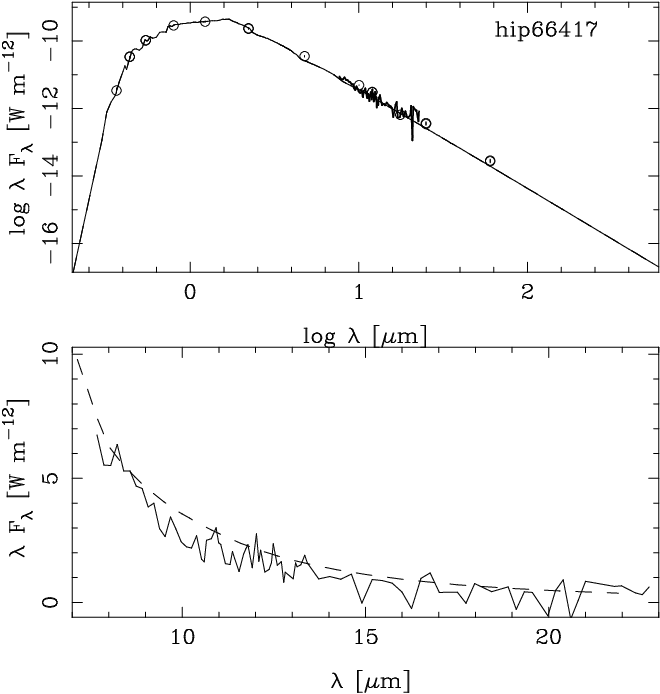}} 
\end{minipage}
\begin{minipage}{0.42\textwidth}
\resizebox{\hsize}{!}{\includegraphics[angle=-0]{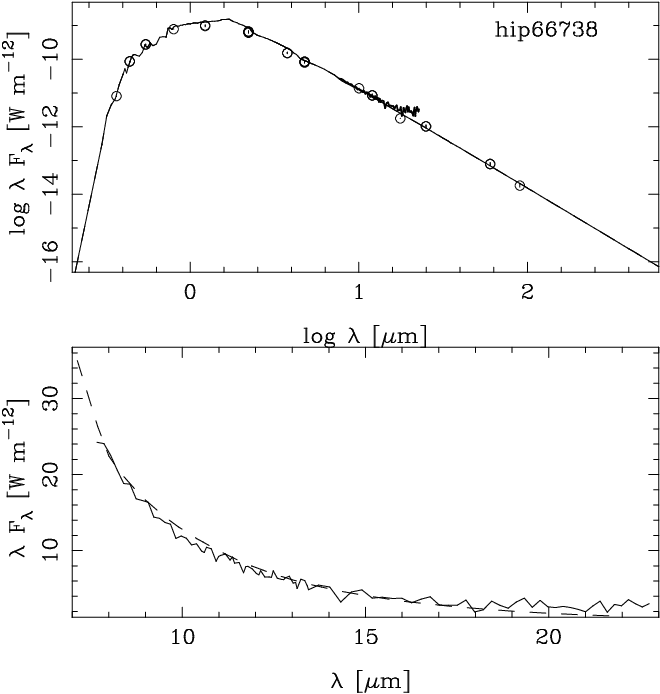}} 
\end{minipage}
 
\caption{Continued. } 

\end{figure*}

\begin{figure*}[H]

\begin{minipage}{0.42\textwidth}
\resizebox{\hsize}{!}{\includegraphics[angle=-0]{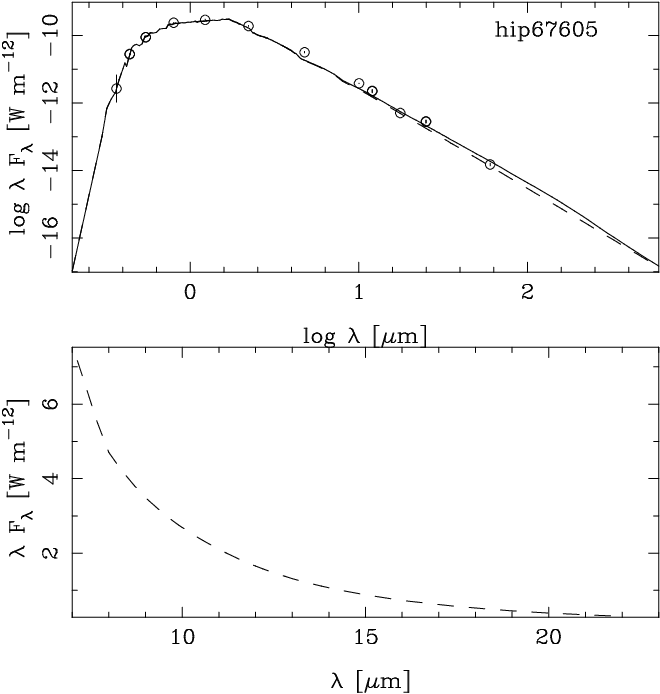}} 
\end{minipage}
\begin{minipage}{0.42\textwidth}
\resizebox{\hsize}{!}{\includegraphics[angle=-0]{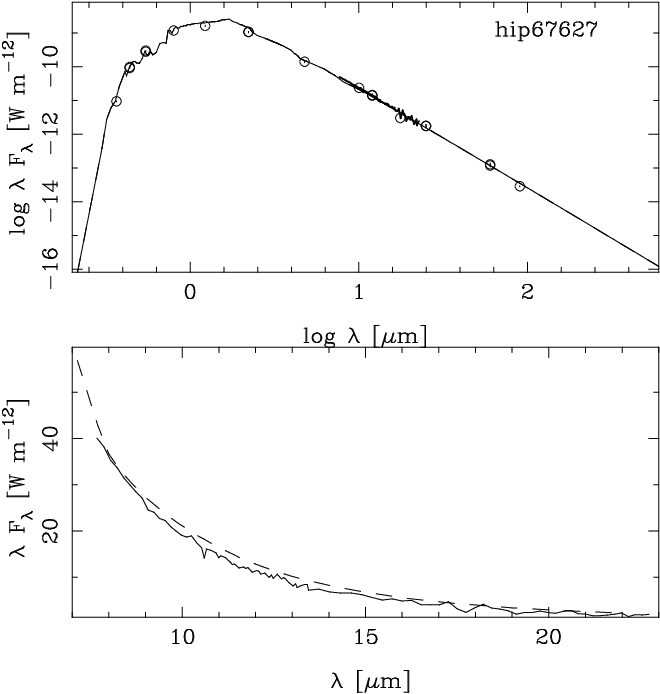}} 
\end{minipage}
 
\begin{minipage}{0.42\textwidth}
\resizebox{\hsize}{!}{\includegraphics[angle=-0]{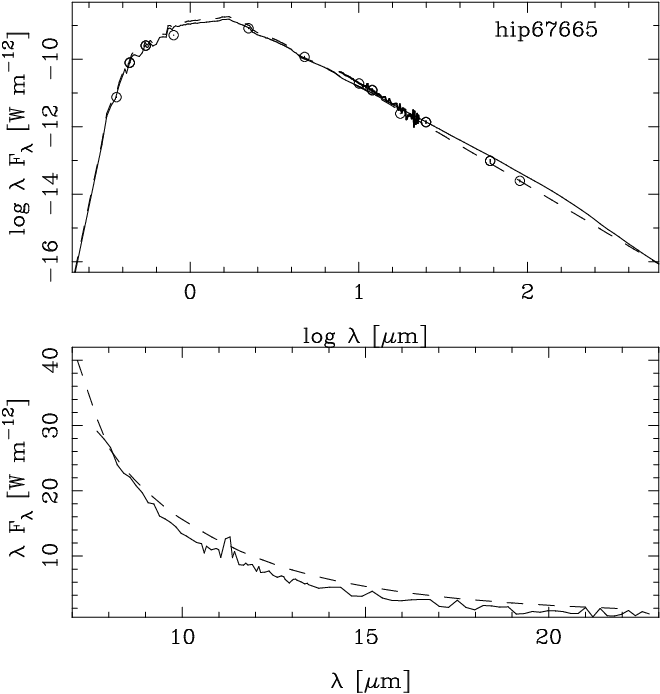}} 
\end{minipage}
\begin{minipage}{0.42\textwidth}
\resizebox{\hsize}{!}{\includegraphics[angle=-0]{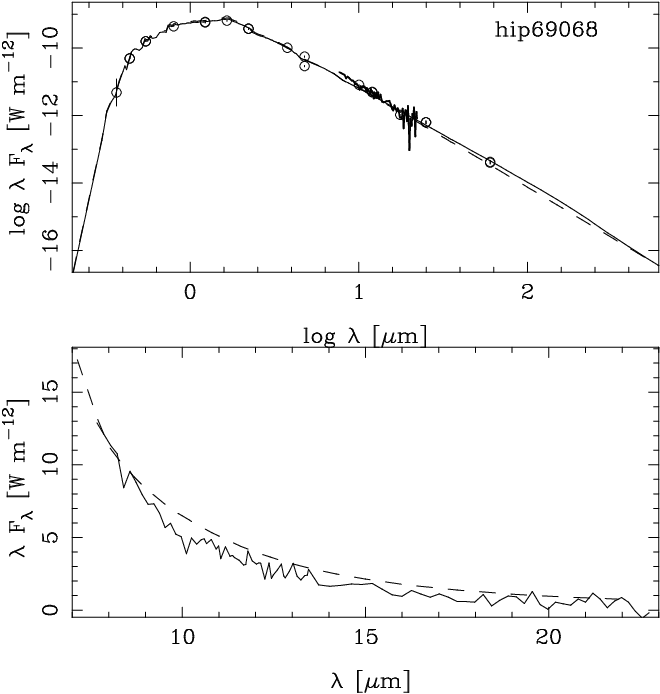}} 
\end{minipage}
 
\begin{minipage}{0.42\textwidth}
\resizebox{\hsize}{!}{\includegraphics[angle=-0]{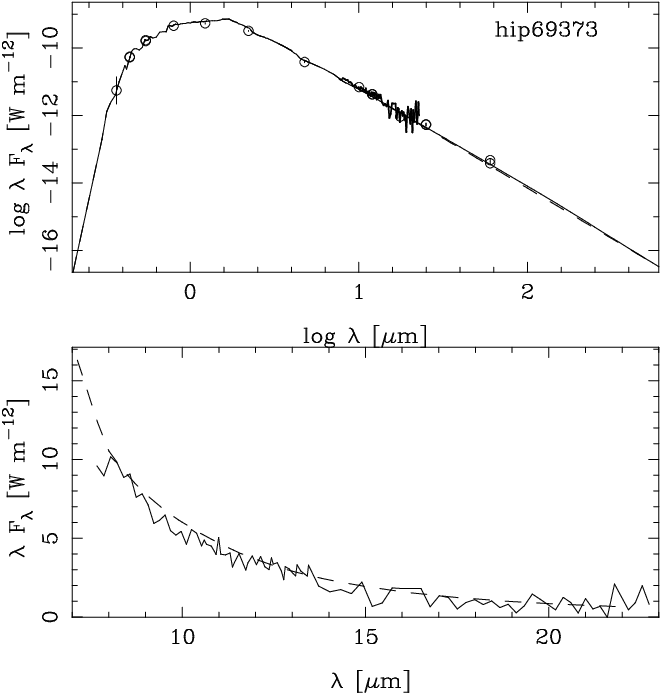}} 
\end{minipage}
\begin{minipage}{0.42\textwidth}
\resizebox{\hsize}{!}{\includegraphics[angle=-0]{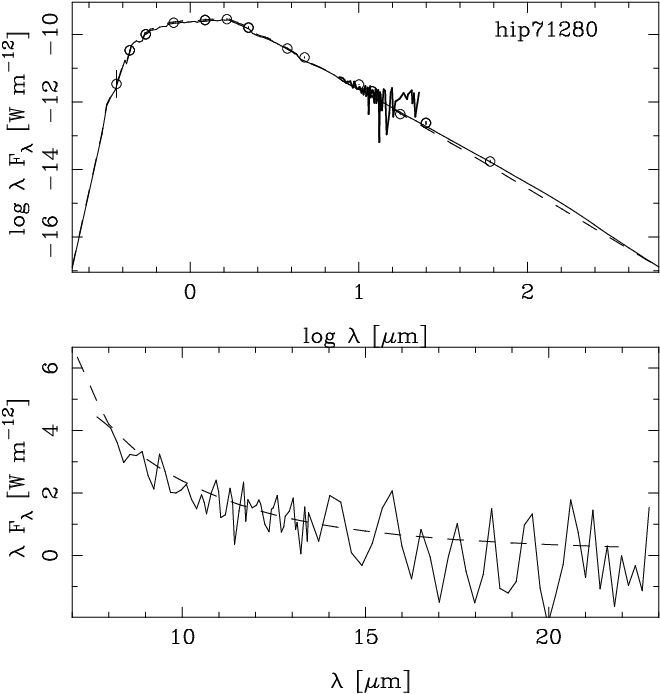}} 
\end{minipage}
 
\caption{Continued. } 

\end{figure*}

\begin{figure*}[H]

\begin{minipage}{0.42\textwidth}
\resizebox{\hsize}{!}{\includegraphics[angle=-0]{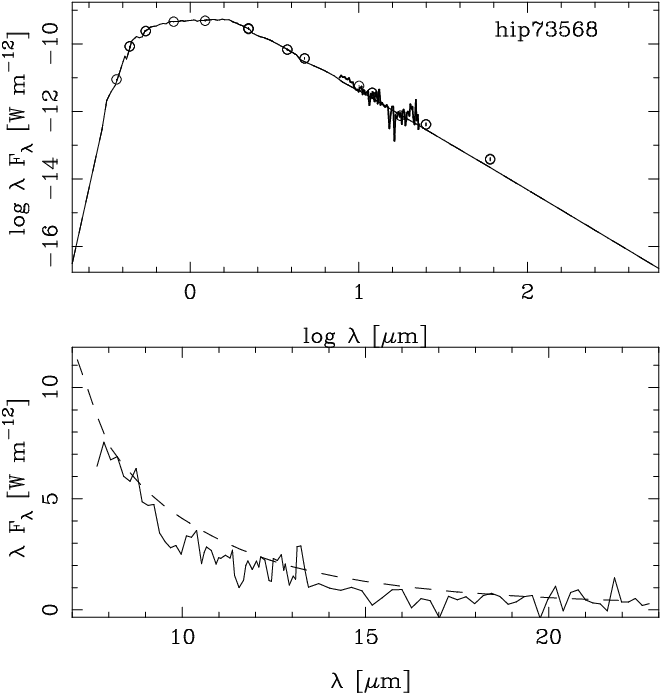}} 
\end{minipage}
\begin{minipage}{0.42\textwidth}
\resizebox{\hsize}{!}{\includegraphics[angle=-0]{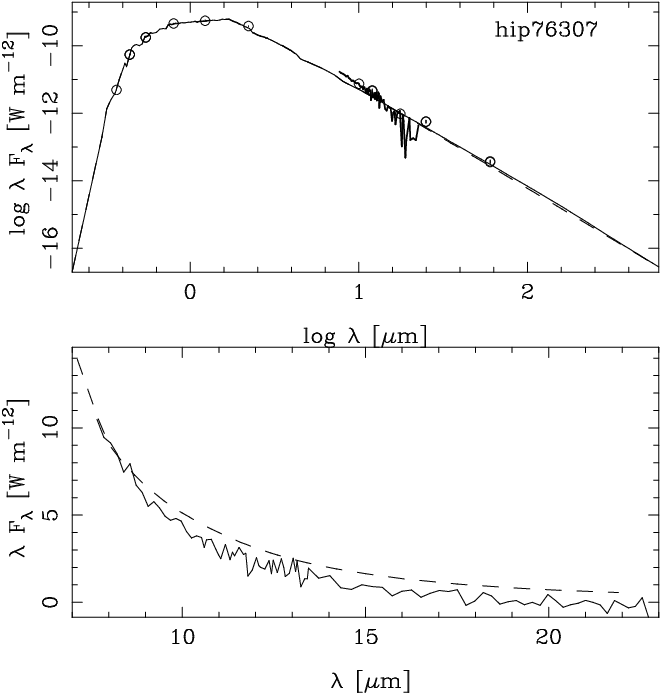}} 
\end{minipage}
 
\begin{minipage}{0.42\textwidth}
\resizebox{\hsize}{!}{\includegraphics[angle=-0]{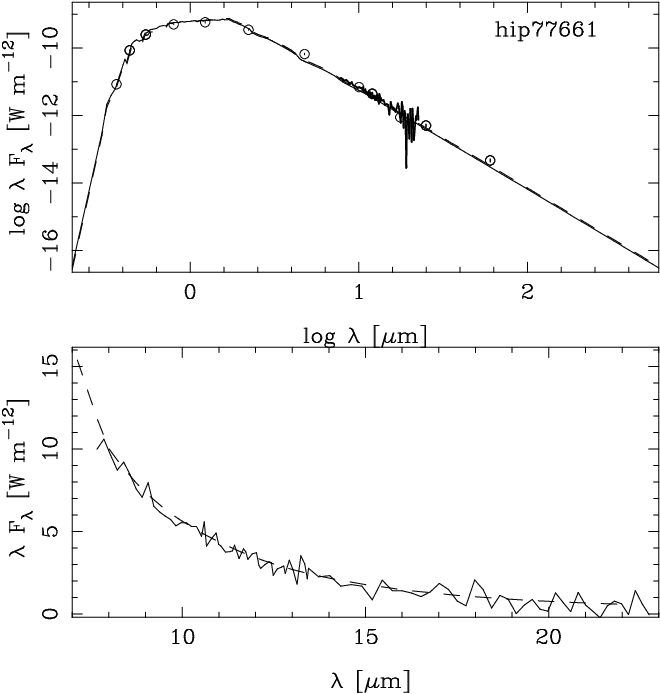}} 
\end{minipage}
\begin{minipage}{0.42\textwidth}
\resizebox{\hsize}{!}{\includegraphics[angle=-0]{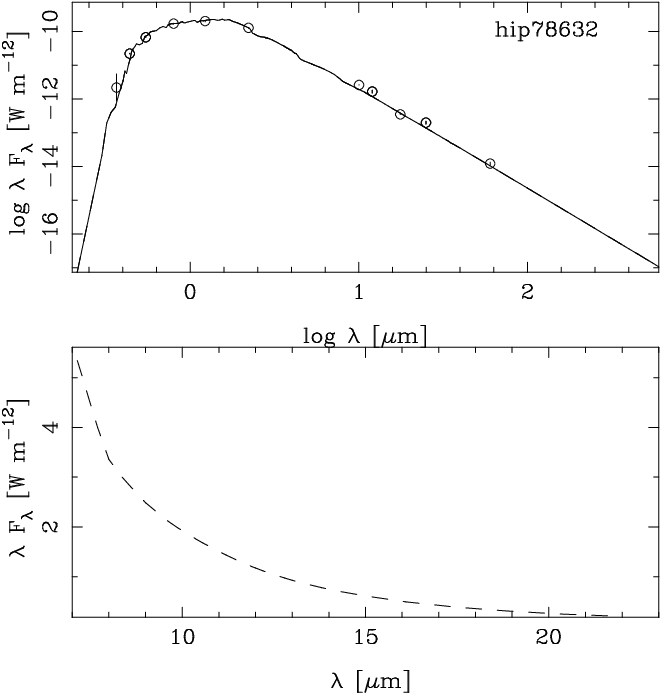}} 
\end{minipage}
 
\begin{minipage}{0.42\textwidth}
\resizebox{\hsize}{!}{\includegraphics[angle=-0]{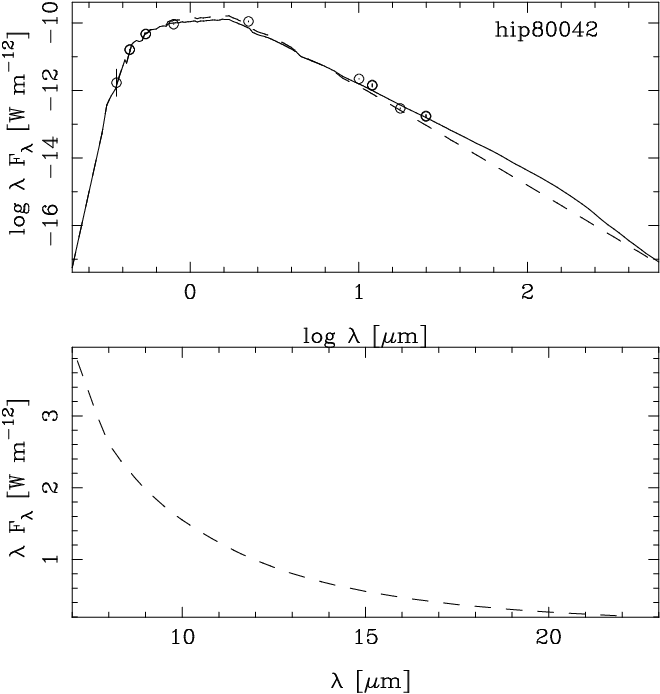}} 
\end{minipage}
\begin{minipage}{0.42\textwidth}
\resizebox{\hsize}{!}{\includegraphics[angle=-0]{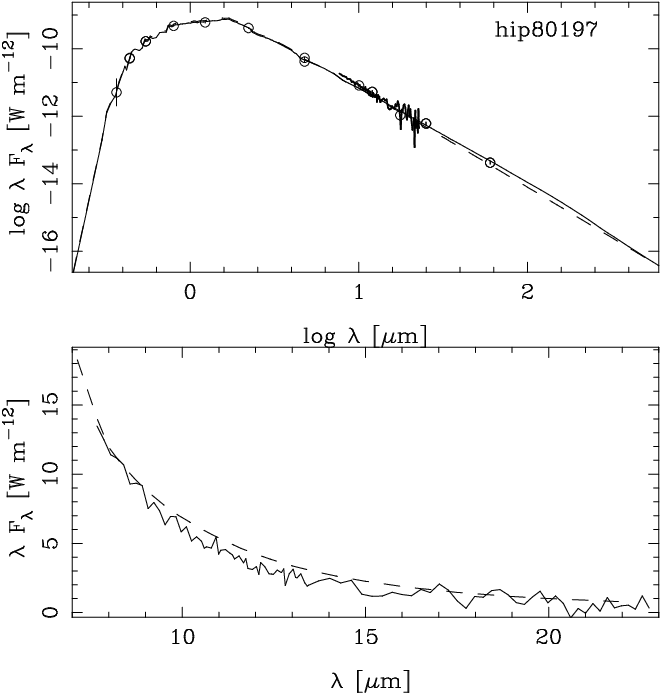}} 
\end{minipage}
 
\caption{Continued. } 

\end{figure*}

\begin{figure*}[H]

\begin{minipage}{0.42\textwidth}
\resizebox{\hsize}{!}{\includegraphics[angle=-0]{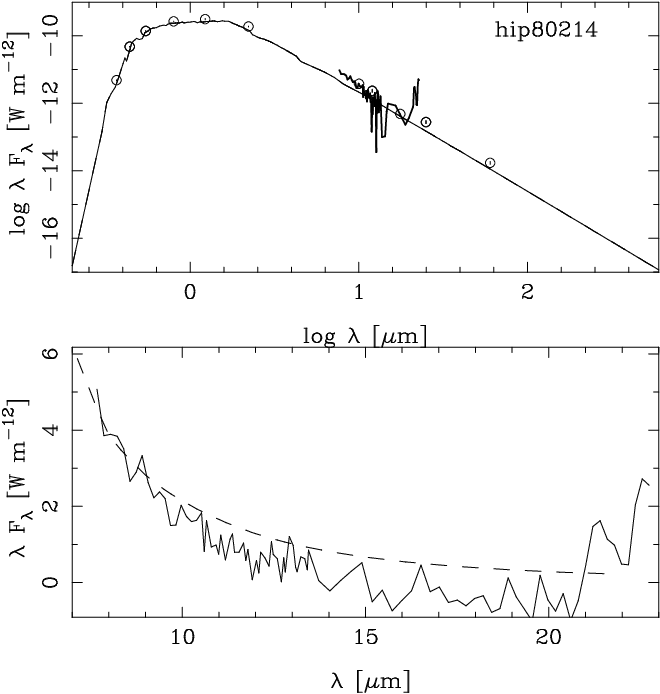}} 
\end{minipage}
\begin{minipage}{0.42\textwidth}
\resizebox{\hsize}{!}{\includegraphics[angle=-0]{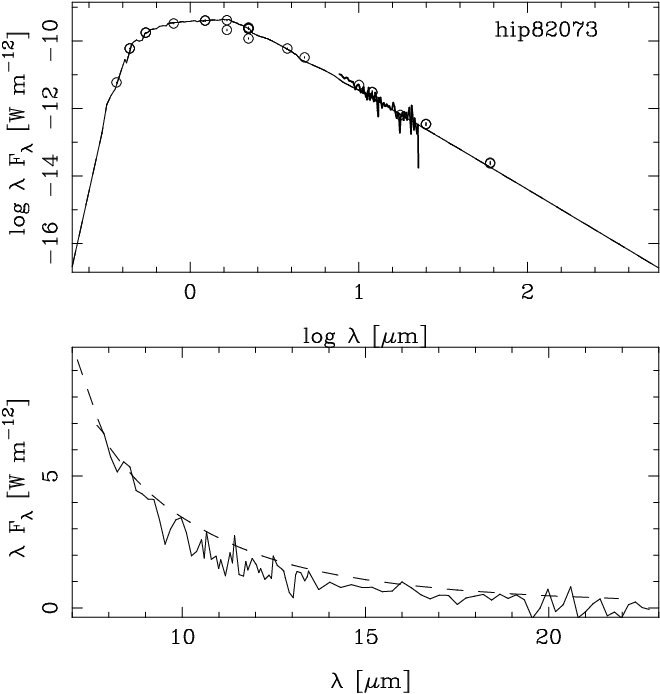}} 
\end{minipage}
 
\begin{minipage}{0.42\textwidth}
\resizebox{\hsize}{!}{\includegraphics[angle=-0]{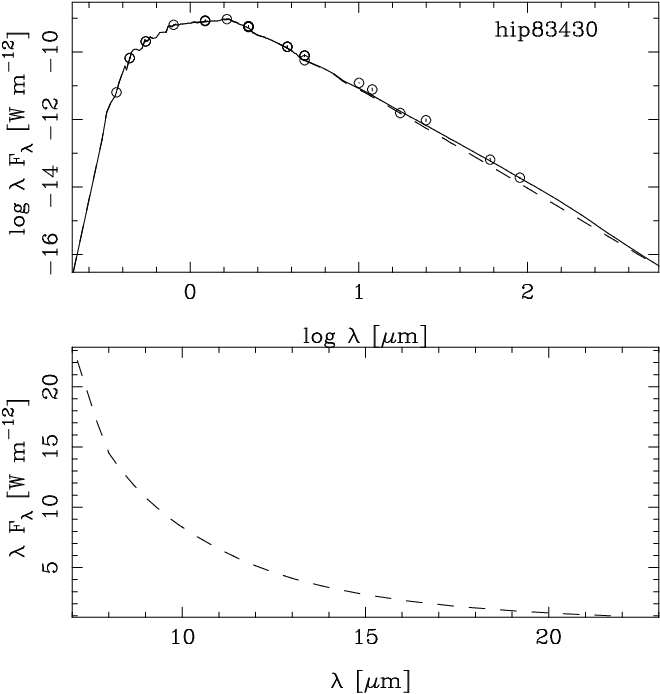}} 
\end{minipage}
\begin{minipage}{0.42\textwidth}
\resizebox{\hsize}{!}{\includegraphics[angle=-0]{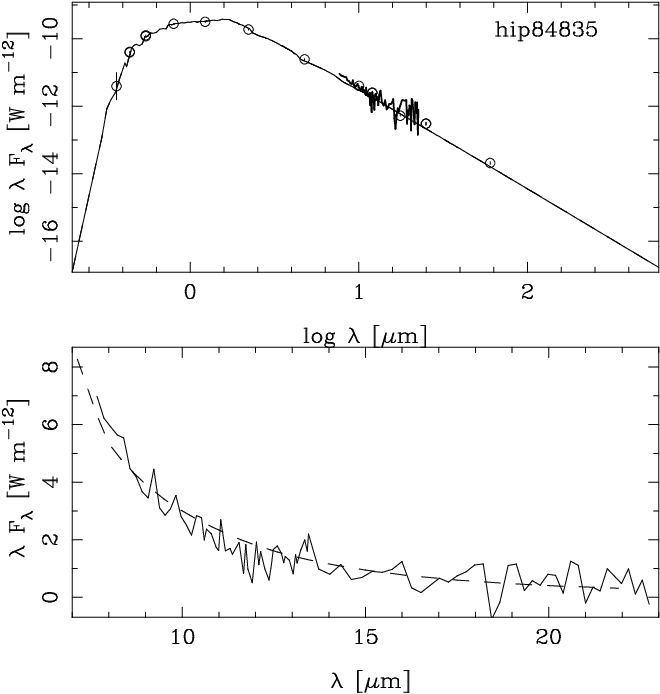}} 
\end{minipage}
 
\begin{minipage}{0.42\textwidth}
\resizebox{\hsize}{!}{\includegraphics[angle=-0]{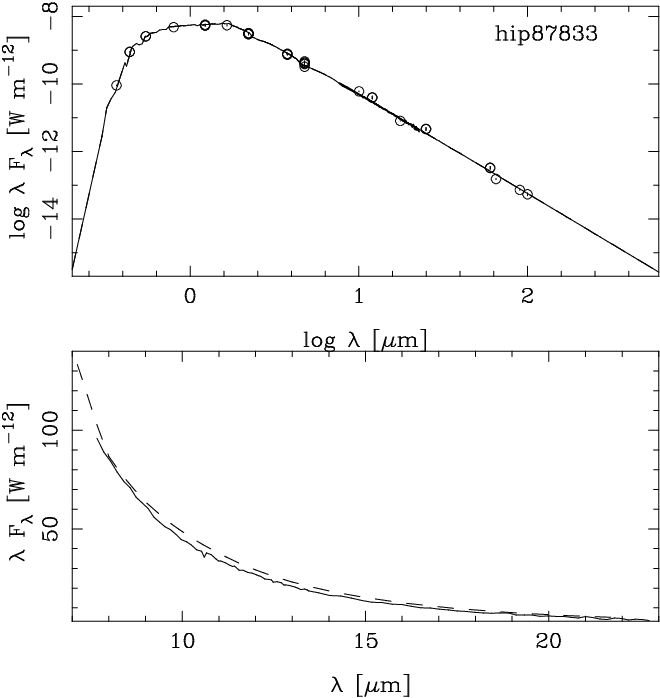}} 
\end{minipage}
\begin{minipage}{0.42\textwidth}
\resizebox{\hsize}{!}{\includegraphics[angle=-0]{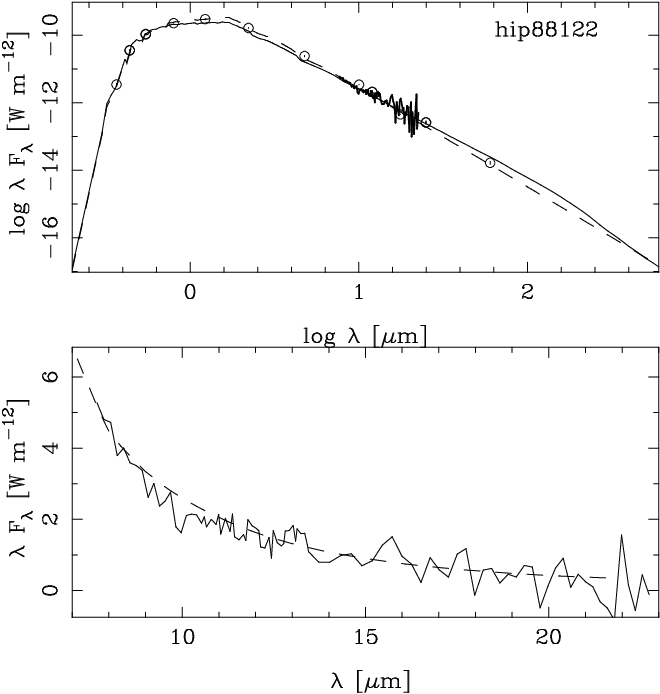}} 
\end{minipage}
 
\caption{Continued. } 

\end{figure*}

\begin{figure*}[H]

\begin{minipage}{0.42\textwidth}
\resizebox{\hsize}{!}{\includegraphics[angle=-0]{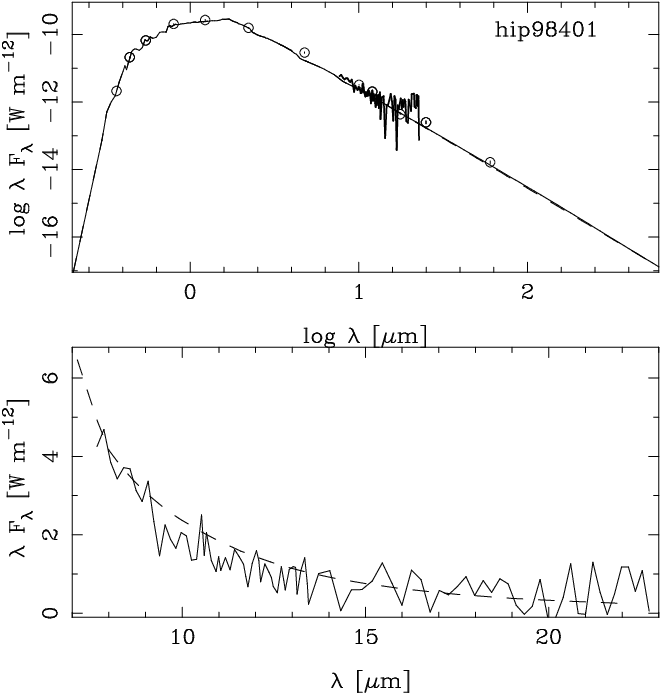}} 
\end{minipage}
\begin{minipage}{0.42\textwidth}
\resizebox{\hsize}{!}{\includegraphics[angle=-0]{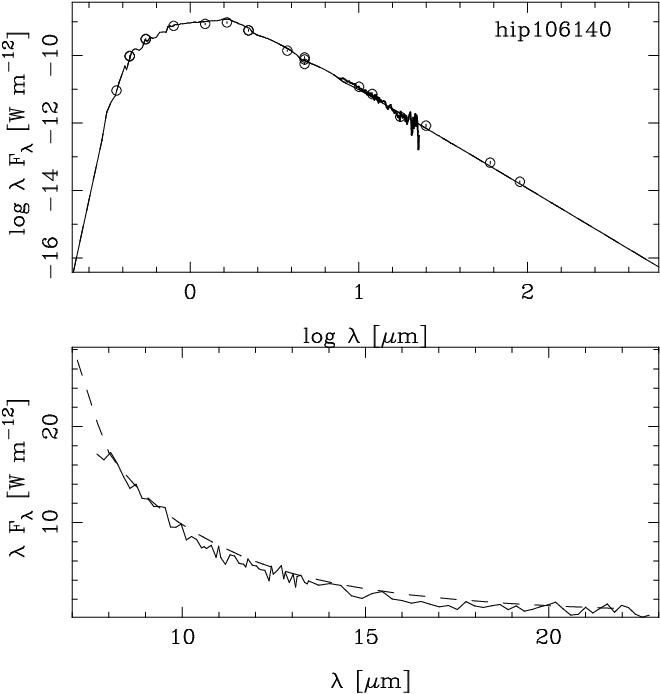}} 
\end{minipage}
 
\begin{minipage}{0.42\textwidth}
\resizebox{\hsize}{!}{\includegraphics[angle=-0]{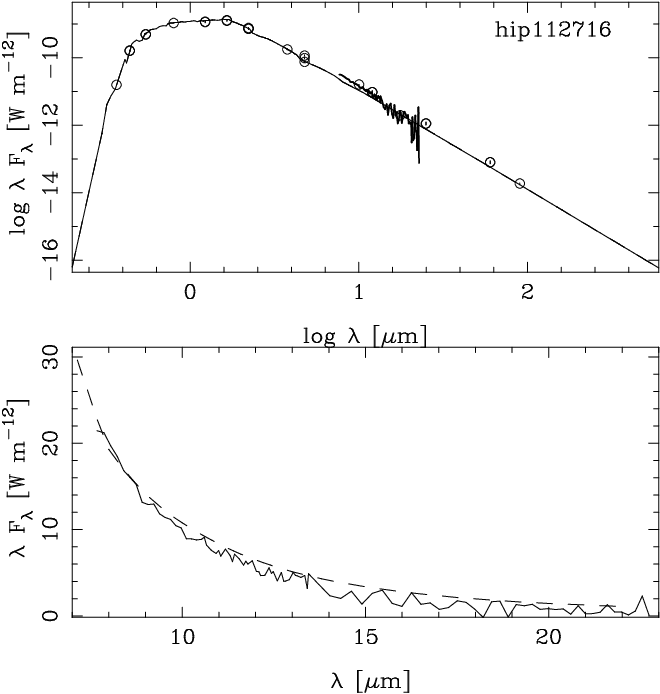}} 
\end{minipage}
\begin{minipage}{0.42\textwidth}
\resizebox{\hsize}{!}{\includegraphics[angle=-0]{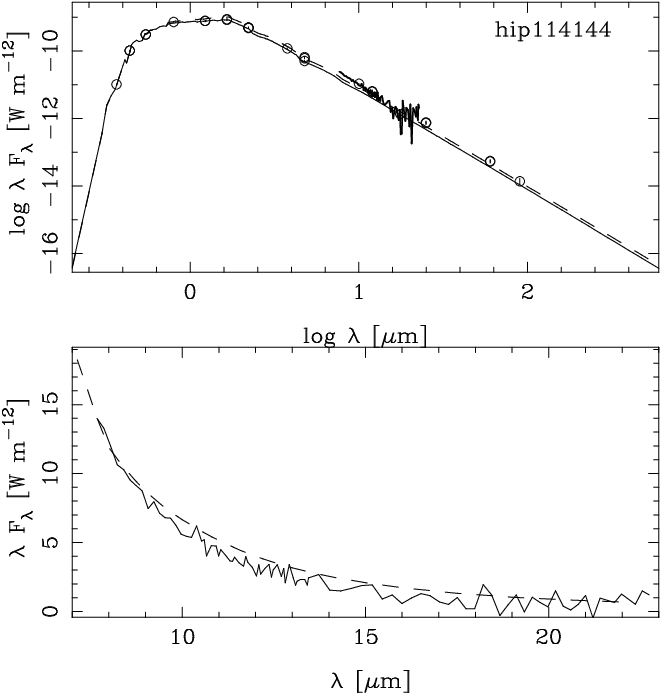}} 
\end{minipage}
 
\begin{minipage}{0.42\textwidth}
\resizebox{\hsize}{!}{\includegraphics[angle=-0]{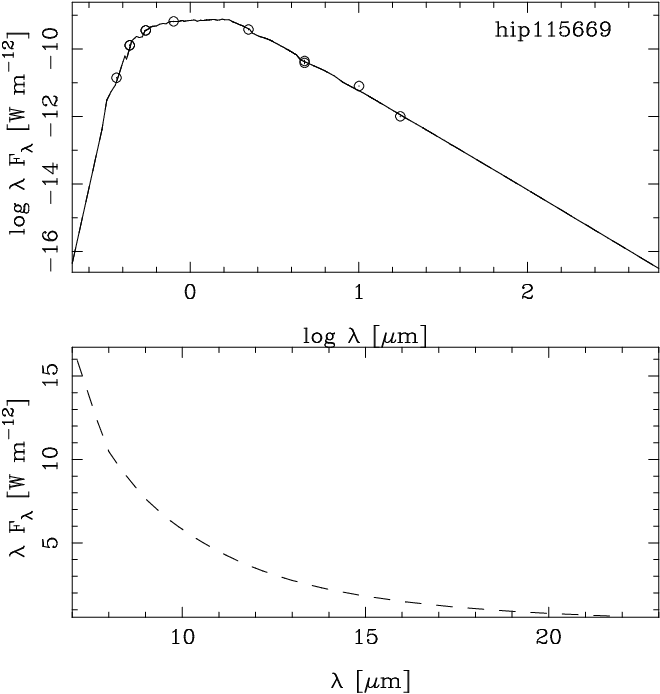}} 
\end{minipage}
\begin{minipage}{0.42\textwidth}
\resizebox{\hsize}{!}{\includegraphics[angle=-0]{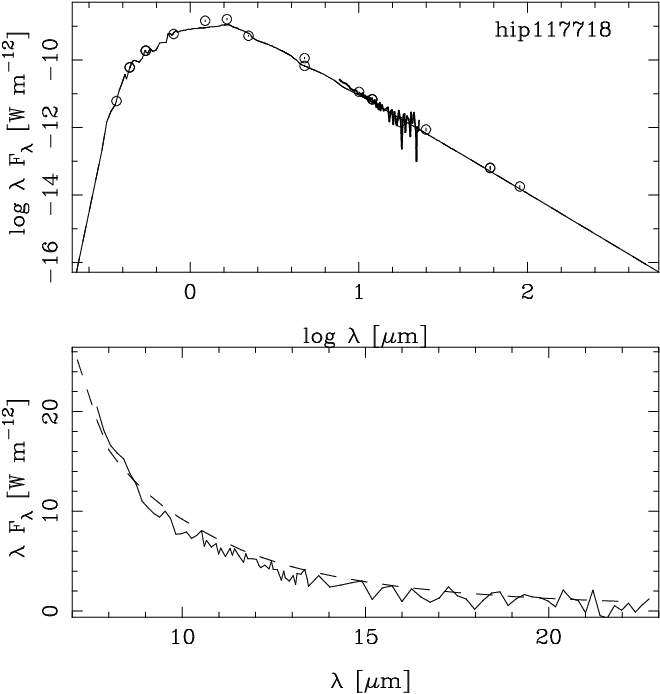}} 
\end{minipage}
 
\caption{Continued. } 

\end{figure*}

\end{appendix}


\begin{thebibliography}{} 

\bibitem[]{} Akritas, M.G., \& Bershady, M.A. 1996, ApJ, 470, 706

\bibitem[]{} Arenou F, Grenon M., \& Gomez A., 1992, A\&A 258, 104

\bibitem[]{} Begemann, B., Dorschner, J., \& Henning, Th. et al. 1997, ApJ, 476, 199

\bibitem[]{} Beichman, C.A., Neugebauer, G., Habing, H. J., Clegg, P. E. \& Chester, T.J., 1985, in IRAS Explamatory Supplement (Pasadena: JPL)

\bibitem[]{} Bell, R.A., \& Gustafsson, B. 1989, MNRAS, 236, 653

\bibitem[]{} Bertelli, G., Girardi, L., Marigo, P., Nasi, E., 2008, A\&A, 484, 815

\bibitem[]{} Blackwell, D.E., Petford, A.D., Arribas, S., Haddock, D.J., \& Shelby, M.J. 1990, A\&A, 232, 396

\bibitem[]{} Bord\'e, P., Coud\'e du Foresto, V., Chagnon, G., \& Perrin, G. 2002, A\&A, 393, 183

\bibitem[]{} Boyer, M.L.,  van Loon, J.Th.,  McDonald, I. et al. 2010, ApJ, 711, L99

\bibitem[]{} Boyer, M.L., McDonald, I., van Loon, J.Th. et al. 2009, ApJ, 705, 746 

\bibitem[]{} Caloi, V., \& d'Antona, F. 2008, ApJ, 673, 847

\bibitem[]{} Catelan, M. 2000, ApJ, 531, 826

\bibitem[]{} Cenarro, A.J., Peletier, R.F., S\'anchez-Bl\'azquez, P. et al. 2007, MNRAS, 374, 664

\bibitem[]{} Clement, C.M.,  Muzzin, A., Dufton, Q. et al. 2001, AJ, 122, 2587

\bibitem[]{} Cohen, M., Walker, R.G., Carter, B. et al. 1999, AJ, 117, 1864

\bibitem[]{} da Silva, L., Girardi, L, Pasquini, L. et al. 2006, A\&A, 458, 609

\bibitem[]{} Drimmel, R., Cabrera-Lavers, A., \&  L\'opez-Corredoira, M. 2003, A\&A, 409, 205

\bibitem[]{} Dumm, T., \& Schild, H. 1998, NewA, 3, 137

\bibitem[]{} Elitzur, M., \& Ivezi\'c, \v Z 2001, MNRAS, 327, 403 

\bibitem[]{} ESA, 1997, The Hipparcos Catalogue, ESA SP-1200 (viZier catalog I/239)

\bibitem[]{} Fern\'andez-Villaca\~nas, J.L., Rego, M., \& Cornide, M. 1990, AJ, 99, 1961

\bibitem[]{} Gezari D.Y., Pitts P.S., Schmitz M., 1999, in Catalog of Infrared Observations, Edition 5 (viZier catalog II/225)

\bibitem[]{} Groenewegen, M.A.T. 2008, A\&A, 488, 935

\bibitem[]{} Groenewegen, M.A.T., Sloan, G.C., Soszy\'nski, I., \& Petersen, E.A. 2009, A\&A, 506, 1277 

\bibitem[]{} Gustafsson, B., Edvardsson, B., Eriksson, K. et al.  2008, A\&A, 486, 951

\bibitem[]{} Habing, H.J., \& Olofsson, H. 2003, in Asymptotic Giant Branch Stars (New York: Springer-Verlag)

\bibitem[]{} Ishihara D., Onaka T., Kataza H., et al. 2010, A\&A, 514, A1 (viZier catalog II/297)  

\bibitem[]{} Ivezi\'c, \v Z., Nenkova, M., \& Elitzur, M. 1999, DUSTY user manual, University of Kentucky internal report

\bibitem[]{} Koen, C. \& Eyer, L. 2002, MNRAS, 331, 45 

\bibitem[]{} Koen, C. \& Laney, D. 2000, MNRAS, 311, 636 

\bibitem[]{} Kudritzki, R. \& Reimers, D. 1978, A\&A, 70, 227

\bibitem[]{} Lebzelter, T., \& Wood, P.R 2005, A\&A, 441, 1117

\bibitem[]{} Marshall, D.J., Robin, A.C., Reyle\', C., Schultheis, M., \& Picaud, S. 2006, A\&A, 453, 635

\bibitem[]{} Massarotti, A., Latham, D.W., Stefanek, R.P. et al. 2008, AJ, 135, 209

\bibitem[]{} Mauas, P.J.D., Cacciari, C., \&  Pasquini, L. 2006, A\&A, 454, 609

\bibitem[]{} Mermilliod, J.C., 1991, in Catalogue of Homogeneous Means in the UBV System (viZier catalog II/168)

\bibitem[]{} McDonald, I., Boyer, M.L., van Loon J.Th., \& Zijlstra A.A. 2011, ApJ, 730, 71 

\bibitem[]{} McDonald, I., Sloan, G.C., Zijlstra, A.A. et al. 2010, ApJ, 717, L92 

\bibitem[]{} McDonald, I., van Loon, J.Th., Decin, L. et al. 2009, MNRAS, 394, 831 

\bibitem[]{} McDonald, I., van Loon, J.Th., Sloan, G.C. et al. 2011, MNRAS, 417, 20  

\bibitem[]{} McWilliam, A. 1990, ApJS, 74, 1075

\bibitem[]{} M\'esz\'aros, Sz., Avrett, E.H., \& Dupree, A.K. 2009, AJ, 138, 615

\bibitem[]{} Miglio, A., Brogaard, K., Stello, D. et al. 2012,  MNRAS, 419, 2077 

\bibitem[]{} Min, M., Hovenier, J.W., \& de Koter, A., 2005, A\&A, 432, 909

\bibitem[]{} Momany, Y., Saviane, I., Smette, A. et al. 2012, A\&A, 537, A2 

\bibitem[]{} Moshir, M., Kopan, G., Conrow, T., et al. 1989, in Explanatory supplement to the IRAS Faint Source Survey (Pasadena: JPL)

\bibitem[]{} Mozurkewich, D., Armstron, J.T., Hindsley, R.B. et al. 2003, AJ, 126, 2502

\bibitem[]{} Niyogi, S.G., Speck, A., \& Onaka, T, 2011, ApJ, 733, 93

\bibitem[]{} Olnon, F.M.,  Raimond, E., Neugebauer, G. et al. 1986, A\&AS, 65, 607

\bibitem[]{} Origlia, L., Rood, R.T., Fabbri, S. et al. 2007, ApJ, 667, L85

\bibitem[]{} Origlia, L., Rood, R.T., Fabbri, S. et al. 2010, ApJ, 718, 522

\bibitem[]{} Planck Collaboration, 2011, A\&A, 537, A7

\bibitem[]{} Pollack, J.B., Hollenbach, D., Beckwith, S.  et al. 1994, ApJ, 421, 615

\bibitem[]{} Reimers, D. 1975, MSRSL, 8, 369

\bibitem[]{} Richichi, A., Percheron, I., \& Davis, J. 2009, MNRAS, 399, 399

\bibitem[]{} Schwarz, G. 1978, Ann. Stat., 6, 461

\bibitem[]{} Tsuji, T. 2000, ApJ, 538, 801

\bibitem[]{} van Leeuwen, F. 2007, in Hipparcos, the new reduction of the raw data (Dordrecht: Springer)

\bibitem[]{} Verhoelst, T., Decin, L., van Malderen R., et al. 2006, A\&A, 447, 311

\bibitem[]{} Verhoelst, T., Van der Zypen, N., Hony, S., Decin, L., Cami, J., \& Eriksson, K. 2009, A\&A, 498, 127

\bibitem[]{} Vieytes, M., Mauas, P., Cacciari, C., Origlia, L., \& Pancino, E. 2011, A\&A, 526, A4

\bibitem[]{} Volk, K., \& Cohen, M., 1989, AJ, 98, 931

\bibitem[]{} Yamamura, I., Makiuti, S., Ikeda, N., et al. 
2010, ISAS/JAXA (viZier catalog II/298) 


\end{thebibliography}
\end{document}